%% file: main_file.tex
\documentclass[sigconf,screen,10pt,table,usenames,dvipsnames,svgnames,x11names]{acmart}
\acmConference[Middleware '22]{23rd ACM/IFIP International Middleware Conference}{November 7--11, 2022}{Quebec, QC, Canada}

\usepackage[utf8]{inputenc}
\usepackage[abbreviations]{foreign}  % For \ie, \eg, \etc
\usepackage{tabularx, makecell, booktabs}
\usepackage{multirow}
\usepackage{subcaption}
\usepackage{graphicx}
\usepackage{xifthen}
\usepackage{hyperref}
\usepackage{pifont}
\usepackage[justification=centering]{caption}
\usepackage{enumitem}
\usepackage{geometry}
\newbool{showcomments}
\setbool{showcomments}{true}
\ifthenelse{\boolean{showcomments}}
{ \newcommand{\mynote}[3]{
    \fbox{\bfseries\sffamily\scriptsize#1}
    {\small$\blacktriangleright$\textsf{\emph{\color{#3}{#2}}}$\blacktriangleleft$}}}
{ \newcommand{\mynote}[3]{}}
\newcommand{\sys}{\textsc{GradSec}\xspace}

\acmYear{2022}\copyrightyear{2022}
\acmConference[Middleware '22]{23rd ACM/IFIP International Middleware Conference}{November 7--11, 2022}{Quebec, QC, Canada}
\acmBooktitle{23rd ACM/IFIP International Middleware Conference (Middleware '22), November 7--11, 2022, Quebec, QC, Canada}
\acmPrice{15.00}
\acmDOI{10.1145/3528535.3565255}
\acmISBN{978-1-4503-9340-9/22/11}
  
\begin{document}

  \title{Shielding Federated Learning Systems against Inference Attacks with ARM TrustZone}
  
%%%%%%%%%%%%%%%%%%%%%%%%%%%%%%%%%%%%%%%%%%%%%%

\author[1]{Aghiles Ait Messaoud}
\affiliation{
    \institution{INSA Lyon, LIRIS}
    \city{Lyon}
    \country{France}
}
\additionalaffiliation{%
    \institution{Ecole nationale Supérieure d'Informatique, Algiers}
    \city{Algiers}
    \country{Algeria}
}
\email{aghiles.ait-messaoud@insa-lyon.fr}

\author[2]{Sonia Ben Mokhtar}
\affiliation{
    \institution{INSA Lyon, LIRIS, CNRS}
    \city{Lyon}
    \country{France}
}
\email{sonia.benmokhtar@insa-lyon.fr}

\author[3]{Vlad Nitu}
\affiliation{
    \institution{Integrate.ai}
    %\footnotemark\thanks{Work done while the author was at CNRS, Lyon, France}}
    \city{Toronto}
    \country{Canada}
}
\additionalaffiliation{
    \institution{INSA Lyon, LIRIS, CNRS}
    \city{Lyon}
    \country{France}
}
\email{vlad.nitu@integrate.ai}

\author[4]{Valerio Schiavoni}
\affiliation{
    \institution{University of Neuchâtel}
    \city{Neuchâtel}
    \country{Switzerland}
}
\email{valerio.schiavoni@unine.ch}

\renewcommand{\shortauthors}{Ait Messaoud, et al.}

%%%%%%%%%%%%%%%%%%%%%%%%%%%%%%%%%%%%%%%%%%%%%

%\pagestyle{plain}  
%\pagenumbering{arabic}

\input{abstract}
 



\maketitle

\input{introduction}

\input{problem}

\input{Background}

\input{Threat_model}
\input{Overview}

\input{Sources_of_gradients_leakage}
\input{system}
\input{Evaluation}
\input{relwork}
\input{conclusion}

\begin{acks}
    This publication incorporates results from the VEDLIoT
    project, which received funding from the European Union's
    Horizon 2020 research and innovation programme under grant
    agreement No 957197.
\end{acks}

{\small
\bibliographystyle{ACM-Reference-Format}
\bibliography{main_file}}

\end{document}

%% file: abstract.tex
\begin{abstract}
Federated Learning (FL) opens new perspectives for training machine learning models while keeping personal data on the users premises. Specifically, in FL, models are trained on the users' devices and only model updates (\ie, gradients) are sent to a central server for aggregation purposes. 
However, the long list of inference attacks that leak private data from gradients, published in the recent years, have emphasized the need of devising effective protection mechanisms to incentivize the adoption of FL at scale. 
While there exist solutions to mitigate these attacks on the server side, little has been done to protect users from attacks performed on the client side. 
In this context, the use of Trusted Execution Environments (TEEs) on the client side are among the most proposing solutions.
However, existing frameworks (\eg, DarkneTZ) require statically putting a large portion of the machine learning model into the TEE to effectively protect against complex attacks or a combination of attacks. 
We present \sys, a solution that allows protecting in a TEE only sensitive layers of a machine learning model, either statically or dynamically, hence reducing both the Trusted Computing Base (TCB) size and the overall training time by up to $30\%$ and $56\%$, respectively compared to state-of-the-art competitors.

\end{abstract}
%\keywords{keywords, keywords}

%% file: introduction.tex
\section{Introduction}
Federated learning (FL) is a distributed machine learning (ML) approach with increasing adoption from academia and industry~\cite{bonawitz2019towards}.
The main advantage of this approach is to train ML models while keeping private user data (photos, vocal recording, purchase data, \etc) under the control of their owners, \eg, end users. 
In a nutshell, a server distributes a global model to several clients.
In turn, they train such model locally with their data, and send back the model gradients (\ie, updates) to the server. 
The latter aggregates the received gradients and iterates by sending the resulting global model to the clients, until a given accuracy is reached. 
However, the long list of recent privacy attacks (\eg,~\cite{zhu2019deep,nasr2019comprehensive,melis2019exploiting}) demonstrates that sharing model gradients in the context of FL training constitutes a threat to clients' privacy.
Indeed, model gradients may leak sensitive information enabling, for instance, the reconstruction of raw data samples~\cite{zhu2019deep, yin2021see} or the learning of hidden features about the data of participating users~\cite{melis2019exploiting} (\eg, gender, race, \etc). 
Such attacks threaten the main motivation behind using FL, \eg the preservation of users' privacy.
Hence, it is paramount to reduce the impact of such attacks, in particular by securing the model gradients and their computation. 
Many software solutions exist to secure the gradients on the server-side.
For instance, secure aggregation~\cite{bonawitz2016practical} forces the server to only observe the aggregated gradients instead of the individual ones. Differential-privacy~\cite{wei2020federated,dwork2008differential} (DP) adds noise to the client gradients before sending them to the server. Finally, homomorphic encryption~\cite{fang2021privacy} leverages aggregation on homomorphically encrypted gradients. 
The previous techniques mask the individual raw gradients to the server, to prevent launching privacy attacks from it. 
However, despite their proven efficiency, these methods do not prevent privacy attacks launched by compromised or malicious clients, and in case of DP, the built model necessarily loses some accuracy.
%Indeed, the latter's devices operate under an OS, that may be infected by some malicious software. 
The scenario of compromised or malicious client is particularly threatening when one deals with millions of lines of code (such as the Android OS powering billions of devices~\cite{web:armincrease}), exposing a large attack surface. 
%Further, with attacks targeting the underlying hardware architecture (\ie Meltdown~\cite{Lipp2018meltdown},  Spectre~\cite{Kocher2018spectre}, \etc) breaking the memory isolation system on all major CPU vendors, these threats become increasingly worrisome.

To guarantee integrity and confidentiality of FL training, one can rely on Trusted Execution Environments~\cite{7345265} (TEEs).
These are recent turn-key solutions that provide program execution with integrity and confidentiality guarantees, available from all major CPU vendors: \eg ARM TrustZone~\cite{pinto2019demystifying}, Intel SGX~\cite{costan2016intel}, AMD SEV~\cite{whitepaper:kaplan2016amd}.
Typically, TEEs can execute secure \emph{enclaves}, shielding read and write access to an application's protected code and data against malicious user applications, compromised OS or system libraries, and even against physical attacks. 
TEEs have been successfully used in a plethora of application domains: design of DRM (Digital Right Management) to restricts access to intellectual property protected by copyright~\cite{jang2015secret}, emulate a secure NFC card for mobile payment~\cite{umar2017trusted}, shield biometric authentication process~\cite{bhargav2014trusted}, \etc. 
In this paper, we rely on ARM TrustZone to mitigate client-side inference attacks in the FL context.
Noteworthy, a large portion of ARM-enabled mobile devices offer native support to TrustZone, making it the most pervasive mobile architecture ranging from sensors, wearables and smartphones to supercomputers~\cite{web:armwebsite}, encompassing a significant part of the FL clients.
One of the challenges in using TrustZone relates to the amount of secure memory typically available to trusted applications (TA).
In fact, TA can only use few MBs of secure memory (in the order of 3-5MB~\cite{amacher2019performance}), as well as inducing additional CPU overhead.

Existing work (\ie, DarkneTZ~\cite{mo2020darknetz}) secures a subset of deep neural network (DNN) layers inside TrustZone enclaves.
However, DarkneTZ lacks support to secure non-successive layers of the underlying DNN model. 
Our experiments (see \S\ref{sec:eval}) prove that such feature is required to protect simultaneously against two privacy attacks targeting separate elements of a model (\ie, the convolutional and the dense parts). 

We present \sys, a framework that leverages TrustZone to secure, at training time, weights or filters of dense~\cite{web:denselayer} or convolutional~\cite{web:convolutionallayer} layers as well as their inputs, outputs and associated computations, by shielding the gradients computation inside enclaves and from external attackers.
\sys can shield non-successive layers, heavily reducing the TEE-related overhead. 

\sys can also change the protected layers across FL cycles, based on a statically fixed probability distribution, to offer a horizontal protection to a model, by going through all subset of layers, without having to secure them all at a given point of time.
In a nutshell, \sys supports two execution modes: \emph{(1)} static: a subset of DNN layers (\ie, one or two separate slices, where a slice is defined as a set of successive layers), to be protected on client-side, is fixed in advance; and \emph{(2)} dynamic: the protected layers change over the FL cycle through a sliding window. 

We implemented \sys on top of OP-TEE and deployed on a Raspberry Pi 3B+, with full support for TrustZone.
We test the security efficiency of \sys against three state-of-the-art privacy attacks: Data-Reconstruction Inference Attack~\cite{zhu2019deep} (DRIA), Membership Inference Attack~\cite{nasr2019comprehensive} (MIA) and Data-Property Inference Attack~\cite{melis2019exploiting} (DPIA). 
We later show that static \sys successfully reduces the impact of DRIA, MIA or both at same time. 
In the latter case, our approach imposes a smaller CPU and TEE memory overhead than DarkneTZ (8\% and 30\% smaller respectively), as it only requires to secure a subset of the layers (\ie, the head and the tail) without the intermediate ones. 
In addition, dynamic \sys reduces the impact of DPIA while maintaining only few protected layers in the TEE enclave in each FL cycle, ensuring 56\% and 8\% less CPU and TEE memory overhead respectively when compared to DarkneTZ. 

% The contributions of this paper can be summarized as follows.
% \sys is the first framework that can store non-successive layers of an ML model inside the TrustZone secure world. 
% We are the first to...\vs{complete the sentence}.
% We contribute ...and \vs{complete the sentence}.

The paper is organized as follows.
\S\ref{sec:prob} defines the target problem.
\S\ref{sec:background} presents a background about Federated Learning (FL), the considered state-of-the-art client-side privacy attacks, the basic architecture of ARM TrustZone, as well as DarkneTZ.
In \S\ref{sec:threat} we present our threat model. 
Then, we introduce the concepts of secure FL with an overview of \sys in \S\ref{sec:deployment}.
We discuss the sources of gradient leakage underlying the design of \sys in \S\ref{sec:leakage}.
The design of \sys is detailed in \S\ref{sec:design}.
%We explain in Section~\vs{where?} the computation theory behind \sys and its design.
In \S\ref{sec:eval} we evaluate \sys using state-of-the art models and against state-of-the-art attacks.
We survey related work in \S\ref{sec:rw}, before concluding in \S\ref{sec:conclusion}.

%% file: problem.tex
\vspace{-10pt}
\section{Problem Statement}
\label{sec:prob}

\begin{table*}[!t]
	\centering
	\setlength{\tabcolsep}{10pt}
	\small
	\begin{center}
	  \rowcolors{1}{gray!10}{gray!0}
      %\begin{tabularx}{|p{5.8cm}|p{2.2cm}|p{2cm}|p{2cm}|p{4cm}|}
      \begin{tabular}{l|rrrr}		  
           \rowcolor{gray!25}
           SOTA Attacks& \textbf{DRIA} & \textbf{MIA} & \textbf{DRIA + MIA} & \textbf{DPIA}\\
		   \midrule
           \rowcolor{gray!1}
           Success measures of the attacks & ImageLoss < $1$ & AUC=0.95 & N/A & AUC=0.99\\
           %\hline
           %\hline
           Required layers in TEE using DarkneTZ & L2 & L5 & L2-L3-L4-L5 & L2-L3-L4-L5 \\
           %\hline
           Required layers in TEE using \sys & L2 & L5 & L2 and L5 & 2 layers in a RR manner\\
           %\hline
           %\hline
           \sys gain in training time & $\equiv$ & $\equiv$ & $-8,3\%$ & $-56.7\%$  \\
          %\hline
           \sys gain in TCB size & $\equiv$ & $\equiv$ & $-30\%$ & $-8\%$ \\
          \hline
      \end{tabular}
	\end{center}
    \caption{\label{table:problem}Success rate of SOTA attacks against LeNet-5 model (line 1); Layers that need to be put in TEEs using both DarkneTZ and \sys to protect the model against attacks launched by a malicious client (lines 2 and 3) and the corresponding gain of \sys compared to DarkneTZ (lines 4 and 5). $\equiv$ indicates similar performance.}
\end{table*}

Protecting FL systems against inference attacks is becoming an increasingly important problem. While server-side inference attacks have attracted a lot of attention in the research community~\cite{bonawitz2016practical,wei2020federated,fang2021privacy}, mitigation mechanisms against client-side inference attacks have been overlooked. 
Specifically, to illustrate the harm a malicious client can perform, we considered three state-of-the-art inference attacks: \emph{(i)} a Data Reconstruction Inference Attack (DRIA)~\cite{zhu2019deep}, which aims a reconstructing a data sample (\eg, an image that was used at training time); \emph{(ii)} a Membership Inference Attack (MIA)~\cite{nasr2019comprehensive}, which aims at inferring whether a given data sample has been used (or not) at training time and \emph{(iii)} a Data Property Inference Attack (DPIA)~\cite{melis2019exploiting}, which aims at inferring sensitive properties (\eg, gender) about the participating users. 
While these attacks were initially devised on the server side, we adapted them to run on a malicious client. 
Specifically, we ran the above attacks on clients participating in an FL training process for the LeNet-5 machine learning model~\cite{lecun1998gradient}, an image classification model trained using the CIFAR-100 dataset. 
Further details on the experimental setup of this experiment as well as further details on the attacks can be found in \S\ref{sec:evaluation} and \S\ref{ssec:fl-attacks-client}, respectively. 
Results depicted in the first line of Table~\ref{table:problem} show that a malicious client successfully manages to run the above attacks. 

An effective way to protect FL clients from the above attacks is to use TEEs on the client side, as previously demonstrated by DarkneTZ~\cite{mo2020darknetz} a system further described in Section~\ref{ssec:darknet}. 
From line 2 of Table~\ref{table:problem}, we see that DarkneTZ successfully protects against DRIA and MIA as putting only one layer inside the TEE makes the attack unsuccessful.
However, as soon as two attacks are considered at once (DRIA + MIA) or a more complex attack is considered (DPIA), DarkneTZ requires putting four out of the five layers of the model inside the TEE, which incurs an important overhead. 
In this paper, we propose a solution that allows protecting non-successive layers inside the TEE (static \sys) as well as a solution for dynamically putting layers inside the TEE in a round-robin (RR) manner (dynamic \sys). 
As shown in the two last lines of Table~\ref{table:problem}, our solution brings an important performance improvement over DarkneTZ, \ie up to $30\%$ gain in terms of TCB size and up to $56\%$ gain in terms of model training time.

%% file: Background.tex
%!TEX root = main_file.tex
\section{Background}\label{sec:background}
This section provides background on federated learning (\S\ref{ssec:fl}), client-side privacy attacks against which \sys protects (\S\ref{ssec:fl-attacks-client}), more details regarding ARM TrustZone, the TEE we have chosen to prototype \sys (\S\ref{ssec:tz}), and finally a description of DarkneTZ, the closest related work (\S\ref{ssec:darknet}).

\subsection{Federated Learning (FL)}\label{ssec:fl}
Federated Learning, initially developed for word prediction on the Google's Gboard keyboard~\cite{yang2018applied}, is a distributed approach to machine learning that trains a model on decentralized data. 
Its principle is to transfer a learning model from a central server to client data (referred to as a ``code-to-data'' approach) rather than the other way around. 
Each client then trains the model locally (on its own device), hence the data never leaves its device. 
Gradients of the resulting model are then sent to a central server, which aggregates them with other clients' gradients. 
This design improves the clients' privacy, a key property feature clearly lacking in centralized machine learning. 
There exist several industrial applications using FL~\cite{li2020review, web:whatisfl}. 
Google uses FL on its mobile apps to improve on-device machine learning models, \eg ``Hey Google'' in Google Assistant to let users issue voice commands~\cite{yang2018applied}. 
On healthcare domain, FL is claimed to be the turn-key solution for making the transition from research to clinical practice by enabling the privacy-preserving learning from confidential medical analysis~\cite{rieke2020future}. 
In manufacturing industries, FL enable multiple companies to train a condition-monitoring system, that monitors a particular condition in machinery (such as vibration, temperature, \etc) to identify changes that could indicate a developing fault, without revealing their respective data and assets~\cite{kanagavelu2021federated} and keep their IP confidential.
However, despite this growing popularity~\cite{web:growingFL}, FL already has been challenged by several privacy attacks, in particular because the shared models can be reverse engineered to identify clients data, or at least some of its features.
%We detail next the three most significant of such attacks.

%\vspace{-14pt}		
\subsection{Client-Side Privacy Attacks in FL}\label{ssec:fl-attacks-client}
Privacy attacks threaten the confidentiality of FL systems in particular for edge devices. 
Solutions exist to restrict gradients model access on the FL server (secure aggregation~\cite{bonawitz2016practical}, Differential-privacy~\cite{wei2020federated,dwork2008differential}, homomorphic encryption~\cite{fang2021privacy}, \etc), preventing the leak of gradients.
%In fact, the FL server is considered as the main third entity which has access to the client gradients, and thus which can lead privacy attacks. 
However, the client-side OS, system libraries and the device itself may also be compromised by memory scraper malwares~\cite{rodriguez2017evolution} that scan the RAM of infected devices, leading to leakage of the gradients as well as full disclosure of client data. 
%Moreover, a malicious client can easily leak the global model, without the need to access specific client gradients. 
Next, we detail how three state-of-the-art privacy attacks operate. % that could be led by a compromised or malicious FL client.

\textbf{Data-Reconstruction Inference Attack (DRIA)}.
This attack~\cite{zhu2019deep} aims at reconstructing an original input data based on the emitted model gradients.
This attack assumes a honest-but-curious attacker running in an FL client device, monitoring the FL training process, particularly the gradients produced, before they are sent to the server. 
The attacker specifically searches for two emitted gradients, produced right after input and those produced due to attacker's random input, respectively. 
Then, through an optimisation algorithm (Adam~\cite{kingma2014adam}, LBFGS~\cite{liu1989limited}, \etc) similar to Gradient Descent, the attacker optimizes (minimizes) the difference between the two gradients which mathematically leads to generate a random data that approximate the targeted input data. 
            
\textbf{Membership Inference Attack (MIA)}~\cite{nasr2019comprehensive} learns whether specific data instances are present in the global model training dataset ($D$). 
The attacker is a malicious FL client with prior knowledge about $D$, \ie it knows some data that are part of $D$ ($D_1 \subset D$) and some which aren't ($D_2 \not\subset D$). 
The attacker builds a binary classifier by training an attack model on the FL model' gradients \emph{wrt.}  $D_1$ and $D_2$. 
To infer the membership probability of a data point, the attacker feeds it to the FL model, and computes its corresponding gradients.
The generated gradients will be used as input data to the attack model. 
The latter will output the membership probability of the gradients corresponding to the former data point.

\textbf{Data-property Inference Attack (DPIA).}
The third attack is DPIA~\cite{melis2019exploiting}.
It infers the probability of presence of a private property (\textit{prop}) seen by the FL model during his local training by one of the FL clients.
Like MIA, it assumes a malicious FL client who trains a binary classifier (attack model) on model gradients ($g_{prop}$, $g_{nonprop}$) against attacker's auxiliary data ($b^{adv}_{prop}$, $b^{adv}_{nonprop}$). 
The gradients are computed using different snapshots of the FL model, taken across several FL cycles. 
To infer the probability of presence of \textit{prop} among batches of data used to train the global model during an FL cycle, the attacker computes the difference between two consecutive snapshots of the global model to get the aggregated gradients, and feeds those to the attack model. 
            
    %\subsection{Trusted Execution Environment (TEE)}
    %A trusted execution environment (TEE), as defined by /***insert ref***/, is an isolated processing environment in which applications can be securely executed independently of the rest of the system (OS or hypervisor). According to /***insert ref***/, a TEE provides an attested and isolated execution environment for applications. The first property certifies that all software running in the environment conforms to the behavior intended by their developers. The second property guarantees that it is impossible to read or modify code or data loaded in a space managed by this environment (enclave). This property also applies to software that is usually trusted (such as Normal OS or hypervisors). The combination of these two properties allows to create, theoretically, tamper-proof applications. 
    %TEE environment is implemented in specific processors that use hardware extensions. Currently, we member two main technologies implementing TEEs: TrustZone /***ref***/ on ARM processors and SGX /***ref***/ on Intel processors.
\begin{figure}[!t]
    \centering
    \includegraphics[scale=0.65]{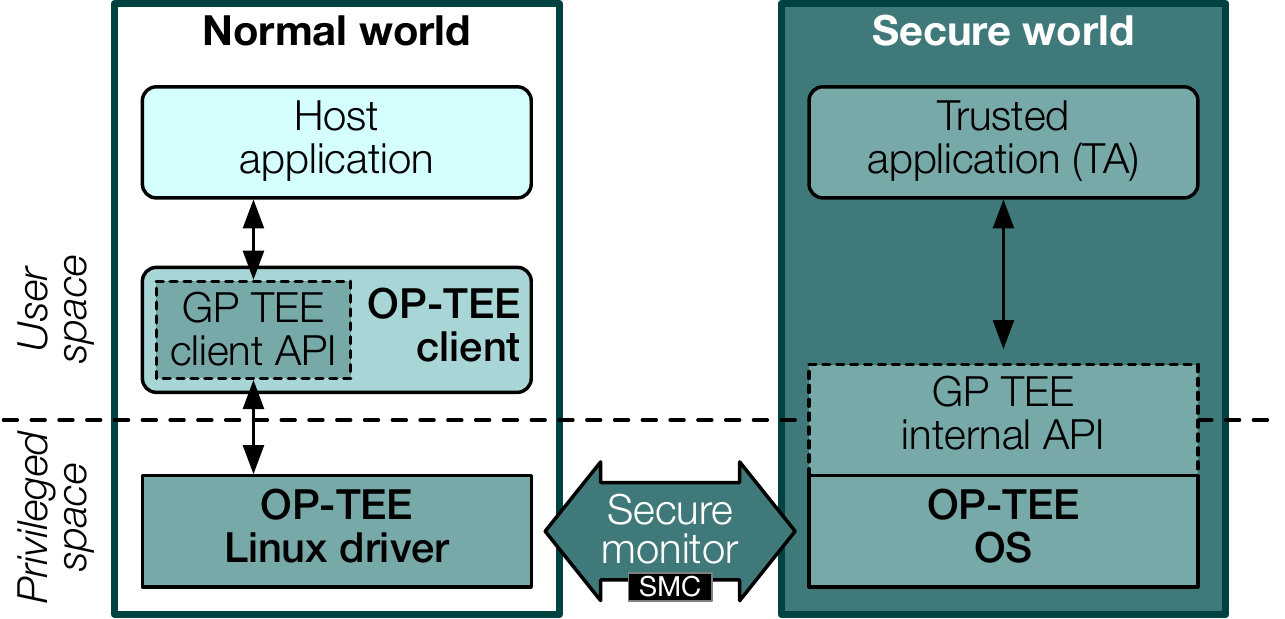}
    \caption{Architecture of REE and TEE using ARM TrustZone. The secure monitor (\texttt{SMC}) allows to switch from the untrusted to the trusted world.)}
    \label{fig:tz}
\end{figure}

\subsection{ARM TrustZone}\label{ssec:tz}
TrustZone~\cite{pinto2019demystifying,amacher2019performance} is the TEE for ARM processors. 
It effectively provides hardware-isolated areas of the processor for sensitive data and code. 
TrustZone splits the execution in two modes (see Figure~\ref{fig:tz}): \emph{(1)} Rich Execution Environment (REE) for \textit{normal} untrusted OS  and \emph{(2)} Trusted Execution Environment (TEE) for \textit{secure} OS. 
The REE is fully managed by the regular OS which executes legacy applications without security guarantees.
The memory, registers and caches of the REE are not protected by any hardware mechanism. 
In contrast, the TEE is managed by the secure OS, which executes trusted applications (TAs) with additional confidentiality and integrity guarantees. 
TAs rely on services, provided by the TEE kernel, to securely access resources (disk, TCP/IP stack, memory...). 
In turn, TAs provide API services for legacy applications as well as other TAs. 
There exist several available implementations for trusted OS with full support for TrustZone. 
Examples include (i) OP-TEE (Open Portable TEE)~\cite{doc:optee}, the Linaro implementation of secure OS for TrustZone; (ii) Kinibi~\cite{web:trustonic}, the TEE OS of Trustonic that uses a microkernel design to enforce isolations between worlds and (iii) Trusty~\cite{web:trusty}, the secure OS implementation of Android that is meant to offer a standard for developping trusted apps for all android devices.
The main limit of TrustZone is its limited footprint size, inlcuding secure memory size (up to 3-5MB), due to its high cost. Such limitation may prevent users from protecting all the layers of a deep neural network (DNN) inside the enclave, requiring to carefully select which layers we need to secure against a specific attack.

\subsection{DarkneTZ}\label{ssec:darknet}
DarkneTZ~\cite{mo2020darknetz} is an open-source DNN framework compatible with TrustZone and the OP-TEE secure OS~\cite{doc:optee}. 
It builds upon Darknet~\cite{darknet}, an open-source neural network framework implemented in C and with support for CUDA. 
DarkneTZ allows users to secure only successive layers of a neural network in a TrustZone TEE enclave. 
Similar to \sys, it attempts to circumvent the concern of the limited memory size of the TEE enclave by asking the user the ability to protect only a portion of contiguous layers of models. 
This solution has proven to be effective in countering MIA by protecting only some of the last layers of the model and also DRIA by protecting some of the first layers.
However, we show in this paper that DarkneTZ is not effective to protect against both attacks simultaneously since, to do so, non successive layers of a machine leaning model need to be protected into the TEE (e.g., the first layers and the last layers), which is not enabled by the framework as stated in the paper~\cite{mo2020darknetz}. Hence, to protect against the above attacks simultaneously, DarkneTZ requires to secure almost all layers of the underlying model, which generates an important overhead. Moreover, DarkneTZ has not been proved to be effective against DPIA, which we also demonstrate in our evaluation section.

%% file: Threat_model.tex
%!TEX root = main_file.tex
\section{Threat Model}\label{sec:threat}
%Our solution aim to secure FL against some of the most important inference attacks (i.e. DRIA, MIA and DPIA) carried on the client side.
We assume an honest-but-curious client-side attacker.
The attacker does not tamper with the FL process and message flow, and it does not attempt to modify the normal message exchanges of the protocol.
Instead, we assume it has physical access to the client machine, where he can execute processes with high/root privileges. 
We further assume that the server side of the FL process is secured using server-grade TEEs (such as Intel SGX)~\cite{sear}, with fewer memory restrictions, or by leveraging  secure aggregation protocols~\cite{bonawitz2017practical} through multi-party computation.

Similar to DarkneTZ~\cite{mo2020darknetz}, we assume feed-forward neural networks~\cite{fine2006feedforward}, such as fully-connected or convolutional neural networks (CNN)~\cite{albawi2017understanding}, also considered by the privacy attacks described in Section~\ref{ssec:fl-attacks-client}.

%% file: Overview.tex
%!TEX root = main_file.tex
\section{Secure Federated Learning}\label{sec:deployment}

\begin{table}[!t]
\centering
\small
\setlength{\tabcolsep}{0pt}
\begin{center}
	\rowcolors{1}{gray!10}{gray!0}
	\begin{tabularx}{\columnwidth}{Xr}
		\rowcolor{gray!50}
		\multicolumn{2}{l}{\textbf{Model Data}} \\
		\rowcolor{gray!25}
  		\multicolumn{2}{l}{\textbf{Model-specific}}\\	 
		\rowcolor{gray!1}
		\textbf{Notation} & \textbf{Designation}\\		 
		$n$ & Number of layers of the model \\     
    	$X$ & Training Batch \\     
    	$Y$ & Matrix of one-hot encoded labels \\     
    	$\hat{Y}$ & Matrix of model predictions \\     
    	$m$ & Batch-size \\   
    	$\lambda$ & Learning rate \\     
    	$Loss$ & Function of model accuracy to optimize \\    	
		\rowcolor{gray!25}
		\multicolumn{2}{l}{\textbf{Specific to layer l}}\\
		\textbf{Notation} & \textbf{Designation}\\
    	%& \multirow{17}{4em}{\rotatebox[origin=c]{90}{\textbf{Specific to a layer $l$}}} 
   		$n_l$ & Number of neurons (if $l$ is Fully-connected) \\   		
   		$K_l$ & Kernel size  (if $l$ is convolutional) \\   		
   		$W_l$ & Matrix of weights (from the kernel if $l$ is convolutional)\\   		
   		$f_l$ & Activation function\\   		
   		$Z_l$ & Output Matrix of layer $l$ $(Z_l=W_l.A_{l-1})$\\   		
   		$A_l$ & Input Matrix of layer $l+1$ $(A_0=X, A_l=f_l(Z_l), A_n=\hat{Y})$\\   		
   		$dW_l$ & Matrix of Gradients w.r.t $W_l$ ($dLoss/dW_l$)\\  		
   		$\delta_l$ & Derivative of the $Loss$ function \emph{wrt} $Z_l$ ($dLoss/dZ_l$) \\
    	\hline  
		\rowcolor{gray!50}  	
		\multicolumn{2}{l}{\textbf{Operations}} \\
		\rowcolor{gray!25}	
		$.$ & Ordinary product\\    
    	$*$ & Hadamard product\\
    	$\otimes$ & Convolutional product\\
    	\hline
    \end{tabularx}
	\end{center}
    \caption{\label{table:notations}Notation and terminology related to FL model.}
\end{table}

\begin{figure}
\centering
	\includegraphics[scale=0.65]{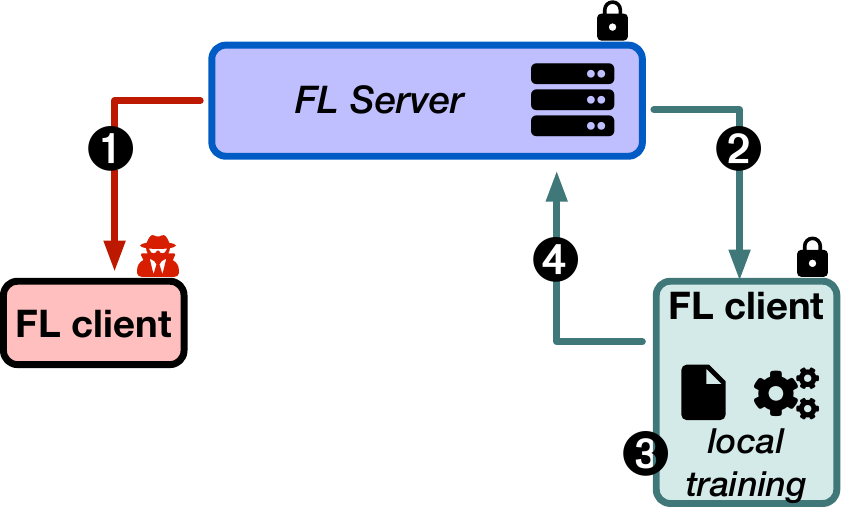}
    \caption{Overview of our approach on secure FL.}
    \label{fig:overview}
	\vspace{-10pt}
\end{figure}

%    \subsection{Overview}
%To understand our approach, we trace the FL process to situate our contribution at a specific stage. 
Figure~\ref{fig:overview} presents an overview of our approach.
We consider a set of clients taking part in the FL training process of a given machine learning model. 
\sys's typical workflow runs as follows.

\noindent \textbf{Selection of FL clients.} To ensure that our approach of securing the local training is effective, the FL server only samples clients with a TEE-compatible device, discarding those without a TEE (Figure~\ref{fig:overview}-\ding{202}).
Therefore, a client interrogation step is required before selecting them for an FL cycle.
The FL server can ensure the trustworthiness of the FL client code leveraging novel remote attestation support, for instance as provided by~\cite{watz}. 
%This interrogation can be done using a TrustZone remote attestation mechanism as in~\cite{pinto2019demystifying}.
        
\noindent \textbf{Transmission of the FL model, hyper parameters and training plan.} Clients receive the FL model, the hyperparameters and the training plan (Figure~\ref{fig:overview}-\ding{203}).
When receiving the FL model, some of its layers should be protected while the others can be left outside the TEE. 
\sys puts the protected layers' weights into the TEE enclave directly using the trusted I/O path (TIOP), as described further in \S\ref{sec:End-to-end security solutions}). 
        
\noindent \textbf{Secure local training.} Each FL client trains his model locally with his own data (see Figure~\ref{fig:overview}-\ding{204}). 
We developed two ways to secure the training. 
In the first one, \ie Static \sys, some layers of the model are permanently protected (\eg, from the first FL cycle until the last) against gradient leakage while the others are not. 
This approach is effective against some state-of-the-art attacks as further discussed in \S\ref{sec:evaluation}. 
In the second approach, Dynamic \sys, the protected layers change along with FL cycles. 
This approach is necessary to protect against more complex attacks as discussed in \S\ref{sec:evaluation}.
In both cases, the data used for training is kept in the storage of the FL client using TrustZone's secure storage~\cite{benedito2021kevlar,hein2015secure} to prevent an external entity from performing a direct leakage of data.
This step implements the main contribution of \sys, with a solution that guarantees a secure local training for the FL clients.
    	
\noindent \textbf{Transmission of the model updates.} Finally, at the end of each FL training cycle, the model gradients of each client are sent to the FL server to be aggregated (Figure~\ref{fig:overview}-\ding{205}).

%% file: Sources_of_gradients_leakage.tex
\section{Sources of Gradients Leakage}\label{sec:leakage}
This section presents the major sources of gradients leakage potentially arising when training a model in a FL context. 
\sys is designed in a way to circumvent these flaws.
Table~\ref{table:notations} summarizes the notations of data and operations related to the FL model training.
We use the popular \textit{Loss} function for multi-class classifiers, \ie the \textit{Categorical Cross-Entropy}~\cite{web:crossentropy}.

In the following, we explain how the gradients of a neural network might leak, a required step to propose the countermeasures contributed by \sys. 
Specifically, we identify two major flaws.
%To secure the gradients of a neural network layer, we studied the different techniques that may leak them. 
%We summarize them in the following two major flaws.

\textbf{$1^{st} Flaw$ : Computing the difference between consecutive states of a model.}
Once the FL client receives the global model from the FL server, it trains it locally with its private data over some local iterations (epochs) and updates the weights of each layer \textit{l}, following the gradient descent formula (SGD) :
                
\begin{equation}
    \begin{array}{ll}
        W_l^{t+1}=W_l^{t}-\lambda dW_l & \mbox{with $t$ the current epoch}
    \end{array}
    \label{eqSGD}
\end{equation}
            
With access to the weights of layer \textit{l}, an attacker can exploit the following simple formula to infer its gradients.
        
\begin{equation}
    dW_l=\frac{W_l^{t+1}-W_l^{t}}{\lambda}
    \label{eqFlaw1}
\end{equation}
            
\noindent Exploiting formula \eqref{eqFlaw1} is our $1^{st} Flaw$.

\textbf{$2^{nd} Flaw$ : Tracking the back propagation computation flow.}
The injection of a batch of data \textit{X} into the model initiates two sequential computations: \emph{(i)} forward propagation and \emph{(ii)} backward (back) propagation. 
The computation of the gradients naturally takes place during the back propagation. 
It consists in a sequence of operations, involving the weights of each layer, that starts from the last layer and propagates backward until the first, in order to compute layer gradients in a descending fashion. 
The formulas for calculating the gradients of a layer \textit{l} during back propagation are, for each type of layer as follows:\\

\noindent For a dense layer :
                    \begin{equation}
                        \label{backpropFC}
                        dW_l=\delta_l . A_{l-1}=\left\{
                            \begin{array}{ll}
                                \frac{1}{m}(\hat{Y}-Y).A_{n-1} & \mbox{if } l=n \\
                                ((W_{l+1}.\delta_{l+1})*f'_l(Z_l)).A_{l-1} & \mbox{if } l<n
                            \end{array}
                        \right.
                    \end{equation}
For a convolutional layer : 
                    \begin{equation}
                        \label{backpropCV}
                        dW_l=\delta_l \otimes A_{l-1} = ((W_{l+1} \otimes \delta_{l+1})*f'_l(Z_l)) \otimes A_{l-1} \mbox{if} l<n
                    \end{equation}
                    For this layer, the \textit{l=n} case is removed because a convolutional layer cannot be at the end of a classifier.\\
                    
\noindent Exploiting formulas \eqref{backpropFC} and \eqref{backpropCV} constitutes our $2^{nd} Flaw$

%% file: system.tex
%!TEX root = main_file.tex
\section{\sys}\label{sec:design}
\begin{figure}[!t]
     \centering     \includegraphics[scale=0.6]{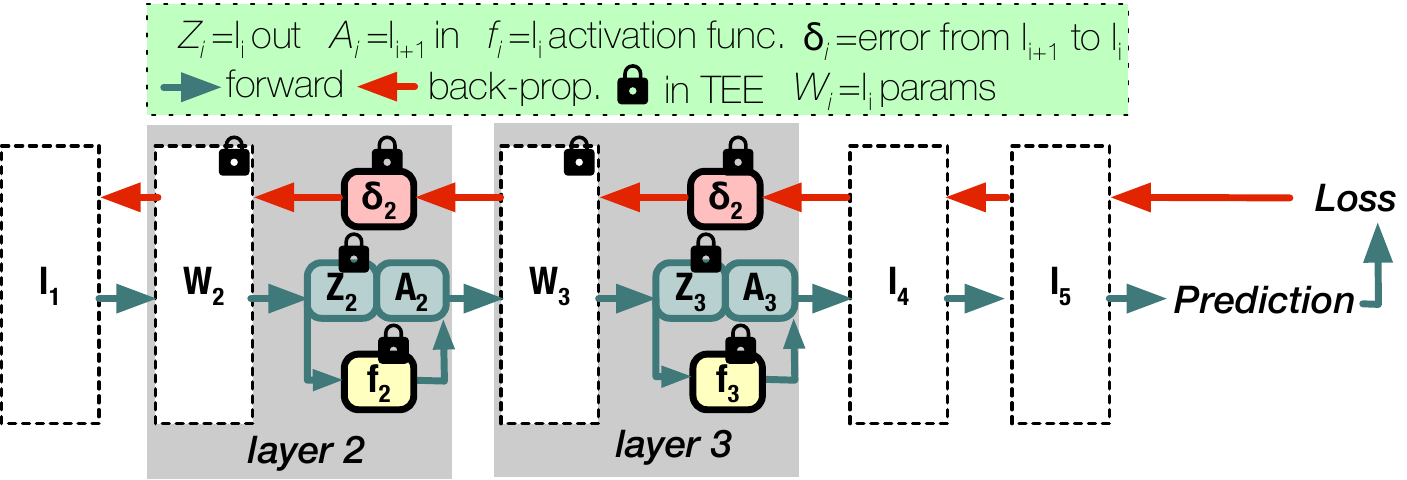}
     \caption{General architecture of $l_2$ and $l_3$ protected layers}
     \label{fig:architecture_of_protected_layer}
 \end{figure}

We present here \sys, a gradient leakage protection scheme, which circumvents the previous flaws by exploiting TEEs. To prevent the attacker from leaking the gradients of layer \textit{l} by any of the previous mathematical formulas, \sys secures in the enclave $W_l$, $Z_l$, $A_{l-1}$ and $\delta_l$ as well as the operations in which they are involved.
Figure~\ref{fig:architecture_of_protected_layer} shows the general architecture of an arbitrary protected layer $l_2$ in a 5-layer neural network. 
\sys works in two modes: static (\S\ref{ssec:static}) and dynamic (\S\ref{ssec:dyn}).
%\vs{/***insert protection schemes***/}

\subsection{Static \sys}\label{ssec:static}
In \emph{static} mode, the FL server fixes in advance a subset of the model layers to be protected in the client-TEE enclave during all the FL cycles. 
This mode is useful when one knows in advance which layers are sensitive to some attacks (\eg, the early/convolutional layers  exploited in the DRIA attack, the tail/dense layers  exploited by the MIA attack, \etc). 
\emph{static} \sys is similar to DarkneTZ~\cite{mo2020darknetz} in its approach to fix, in advance, protected layers.
Yet, \sys has the ability to protect non-successive layers inside the TEE enclave, a subtle yet key difference against DarkneTZ. 
The latter feature is interesting to secure two \emph{distant} layers of a neural network.
For instance,  one could protect layers from the convolutional part that extract meaningful characteristics from the data, and layers from the fully-connected part that usually classifies them. 
This option is fundamental to simultaneously protect against DRIA and MIA attacks, while avoiding to secure the intermediate layers and hence reduces the overall TCB size,  without penalizing security. 
The only parameter required to use \emph{static} \sys is the list of layers to protect.

\subsection{Dynamic \sys}\label{ssec:dyn}
\begin{figure}[!t]
    \centering            \includegraphics[scale=0.7]{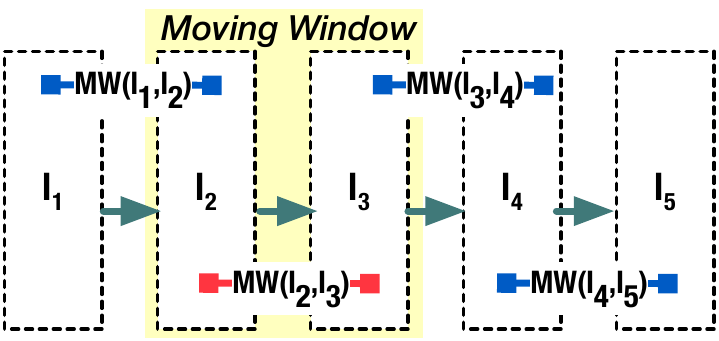}
    \caption{Possible locations of the moving window (MW) in a 5-layers neural network.}
    \label{fig:possible_locations_of_MW}
\end{figure}

With this mode, the layers protected by the TEE, at each client level, change through a moving window (\textit{MW}) over FL cycles. 
The parameters configuring this approach are fixed by the FL server. These are:
\begin{enumerate}[align=parleft,left=0pt..16pt]
	\item {$size_{MW}:$} the number of successive layers to be put in the TEE enclave in each cycle (fixed number for all cycles).
                
	\item {$V_{MW}:$} the probability distribution vector of the \textit{MW} location. It expresses the probability that a given set of successive layers remain on TEE for each FL cycle before MW moves. As an example, Figure~\ref{fig:possible_locations_of_MW} shows four possible locations for an MW of size 2 in a 5-layer neural network. If $V_{MW}=[0.1, 0.3, 0.4, 0.2]$ then the $MW$ will spend $10\%$ of the FL cycles protecting $(l_1,l_2)$, $30\%$ of the FL cycles protecting $(l_2,l_3)$, $40\%$ of the FL cycles protecting $(l_3,l_4)$ and $20\%$ of the FL cycles protecting $(l_4,l_5)$.            
The number of possible locations for an MW (size of $V_{MW}$) in a neural network with $n$ layers is: $n-size_{MW}+1$.
\end{enumerate}
        
The main intuition behind dynamic \sys is to try to protect all the layers of a DNN without putting them all at once inside the TEE, due to its limited size. 
Thus, the MW acts as a sliding TEE region that can only host limited subset of the layers at once. 
Further, since each group of layers covered by the MW may have different sensitivity toward an attack, we offer the ability to customize the protection probability for each group through $V_{MW}$. 
As shown later in our evaluation (\S\ref{sec:eval}), such strategy is more effective than statically protecting layers against DPIA, where the sensitivity toward the attack is unequally distributed among all the layers.

\subsection{End-to-end security solutions}
\label{sec:End-to-end security solutions} 
\noindent\textbf{Trusted I/O path. } 
To safely interact with the device's peripherals, TrustZone enables the reflection of the world state of the processor into the peripherals~\cite{pinto2019demystifying}. Specifically, the client network interface could receive the model weights, related to the protected layers, from the FL server, and safely transfer them in the TEE secure memory throughout a secure channel. 
        
\noindent\textbf{Secure Storage. }
%This TEE feature mandates that it should be possible to store general-purpose data and key material that guarantee confidentiality and integrity of the stored data and the atomicity of the operations that modifies the storage. 
TrustZone's secure storage~\cite{web:securestorage} enables storing general-purpose data and key material while guaranteeing confidentiality and integrity of the latter and the atomicity of the operations that modify them. It leverages a randomly generated File Encryption Key (FEK) for encrypting and decrypting the data stored in block file. The FEK itself is encrypted/decrypted by the Trusted Application Storage Key (TSK) which is derived from the per-device Secure Storage Key (SSK) and the TA's identifier (UUID).
\sys could leverage such functionality to guarantee the confidentiality and integrity of the received FL model, as well as the client data on its device persistent storage, outside of the training time (likely between FL cycles). 
Specifically, in the case of the \sys prototype built on top of the OP-TEE trusted OS, two existing implementations for secure storage~\cite{web:securestorage} exist, namely REE File System or RPMB File System, depending on the underlying hardware support of the client device.
        
\noindent\textbf{Remote attestation. }
Remote attestation (RA) allows remote parties to check the integrity and authenticity of the TEE environment~\cite{kostiainen2011practical}. 
It constitutes a fundamental building block for establishing trust between a TEE and a remote party. 
RA allows the FL server to ensure that the client code is correctly executed in the TEE enclave. 
Despite the lack of native support for RA for TrustZone enclaves, support can be provided by leveraging novel solutions~\cite{watz} or by the incorporation of a hardware chip (\eg Trusted Platform Module) that contains trusted code for measuring the integrity of the TEE kernel and cryptographic keys~\cite{zhao2014providing}.

%% file: Evaluation.tex
%!TEX root = main_file.tex
\label{sec:evaluation}
\section{Evaluation}\label{sec:eval}
This section presents the experimental evaluation of our \sys prototype.
We consider two distinct ML models and real-world datasets. 
To evaluate the performance of \sys, we choose the models and the datasets where each attack performs the best so as to measure the real efficiency of \sys. 
    %We use the FL training model and the dataset with which it works the best : 
We launched DRIA and MIA against the LeNet-5~\cite{lecun1998gradient} (4 convolutional layers and 1 dense layer) and AlexNet~\cite{krizhevsky2012imagenet} (5 convolutional layers and 3 dense layers) models using CIFAR-100~\cite{krizhevsky2009learning}.
We rely on LeNet-5~\cite{lecun1998gradient} using the LFW dataset~\cite{LFWTech} to launch DPIA.
Our threat model consider DRIA and MIA as \emph{single-shot} attacks, \ie they can be performed by an attacker in one FL cycle.
Instead, DPIA is a \emph{long-term} attack, as it needs several FL cycles to collect as many gradients as possible during the model evolution. 
We deploy Dynamic \sys against DPIA, to observe the impact in changing the protected layers along the FL cycles.
Table~\ref{table:configuration} recaps these configurations. 
Table~\ref{table:architecture of models} details the architecture of each model.
    
\begin{table}[!t]
	\centering
	\small
	\setlength{\tabcolsep}{8pt}
	\begin{center}
		\rowcolors{1}{gray!10}{gray!0}
		\begin{tabularx}{\columnwidth}{Xrrr}
       	\rowcolor{gray!50}
         \textbf{Attacks} & \textbf{Models} & \textbf{Datasets} & \textbf{Protection method}\\
         DRIA & LeNet-5 & CIFAR-100 & Static \sys\\
         MIA & AlexNet & CIFAR-100 & Static \sys\\
         DPIA & LeNet-5 & LFW & Dynamic \sys \\
         \hline
    	\end{tabularx}
		\end{center}
    	\caption{Models, datasets and protection method per attack.}
    \label{table:configuration}
\end{table}

\subsection{Evaluation settings} 
%To measure the security level provided by \sys, 
We rely on an existing Python implementation of DRIA~\cite{github:DRIA_github}, MIA~\cite{github:MIA_github} and DPIA~\cite{github:DPIA_github}. 
DRIA rely on the LBFGS~\cite{liu1989limited} optimization algorithm to perform the attack, while MIA and DPIA rely on a dataset of leaked gradients ($D_{grad}$), built by the attacker. 
To mimic the layer-level gradient confidentiality offered by a TEE enclave, we simply delete from $D_{grad}$ all the gradients columns relative to a protected layer since the latter are considered as unavailable for an attacker located in the \textit{normal world}. 
In practice, to dynamically change the protected layers at each FL cycles, we inject the required configuration in $D_{grad}$ only if the moving window \emph{MW} should protect the concerned layer for the given FL cycle. 
%That being said, attacker's $D_{grad}$ becomes a "holey" gradient dataset, where we fill the gaps using the mean of the column.

To measure the performance impact at deployment of \sys, 
We implement and evaluate \sys using Raspberry Pi 3B+, a popular yet representative single-board device, equipped with Broadcom BCM2837B0 (ARM Cortex A53 quad core @ 1.4GHz, 1GB LPDDR2) to mimic an FL client which trains a model. 
We use the latest stable release of OP-TEE~\cite{doc:optee}, a secure OS with TrustZone support. 
We leveraged the latest stable release of DarkneTZ~\cite{mo2020darknetz} as a privacy-preserving deep learning framework to build \sys. 
For static \sys, we extended DarkneTZ  to protect two slices of non-successive layers,  in order to efficiently protect against DRIA and MIA simultaneously. 
For dynamic \sys, we rely on the vanilla DarkneTZ implementation. %since we don't need to secure non-successive layers through the \textit{MW}.

\begin{table}[!t]
	\centering
	\small
	\setlength{\tabcolsep}{3pt}
	\begin{center}
		\rowcolors{1}{gray!10}{gray!0}
		\begin{tabularx}{\columnwidth}{Xlllrr}
     	\rowcolor{gray!50}
     	\multicolumn{6}{c}{\textbf{LeNet-5}}\\
     	\rowcolor{gray!25}
     	\textbf{Layer} & \textbf{Type} & \textbf{\#Filters} & \textbf{FS/S/P} & \textbf{in. size} & \textbf{out. size} \\
		\rowcolor{gray!1}
          L1 & Conv2D & 12 & 5*5/2/0 & 32*32*3 & 16*16*12\\
          L2 & Conv2D & 12 & 5*5/2/2 & 16*16*12 & 8*8*12\\
          L3 & Conv2D & 12 & 5*5/1/2 & 8*8*12 & 8*8*12\\
          L4 & Conv2D & 12 & 5*5/1/2 & 8*8*12 & 8*8*12\\
          L5 & Dense  & /  & /     &  768   & 100\\
		 \rowcolor{gray!50}	
		 \multicolumn{6}{c}{\textbf{AlexNet}}\\
      	 \rowcolor{gray!25}
     	\textbf{Layer} & \textbf{Type} & \textbf{\#Filters} & \textbf{FS/S/P} & \textbf{in. size} & \textbf{out. size} \\
		 \rowcolor{gray!1}
          L1 & Conv2D +MP2 & 64  & 3*3/2/1 & 32*32*3 & 8*8*64\\
          L2 & Conv2D +MP2 & 192  & 3*3/1/1 & 8*8*64 & 4*4*192 \\
          L3 & Conv2D & 384 & 3*3/1/1 & 4*4*192 & 4*4*384\\
          L4 & Conv2D & 256 & 3*3/1/1 & 4*4*384 & 4*4*256\\
          L5 & Conv2D +MP2 & 256 & 3*3/1/1 & 4*4*256 & 2*2*256\\
          L6 & Dense &    /   &   /    & 1024 & 4096 \\
          L7 & Dense &    /   &   /    & 4096 & 4096 \\
          L8 & Dense &    /   &   /    & 4096 & 100 \\
     \end{tabularx}
 	\end{center}
     \caption{Architecture of LeNet-5 and AlexNet. Conv2D: Convolutional layer; MP2: 2*2 MaxPool layer. FS/S/P: filter size/strides, pad.}
     \label{table:architecture of models}
	 \vspace{-20pt}
\end{table}

\subsection{Security Analysis}\label{ssec:security_analysis}
We quantify the performance of DRIA using the \textit{Image Loss} metric, \ie the euclidean distance between the attacker's inferred image and the original FL client image fed to the model. 
We measure the performance of MIA and DPIA using \textit{AUC}, \ie an aggregated measure of the attack model performance considering all the possible classification thresholds. 
It is statistically consistent and more discriminating measure than accuracy~\cite{ling2003auc}. 
An attack model with an AUC of 0.5 is considered as inefficient and performing a random guess regardless the classification threshold. 

\begin{figure}[!t]
    \begin{subfigure}{.5\textwidth}
        \centering
        \includegraphics[scale=0.8]{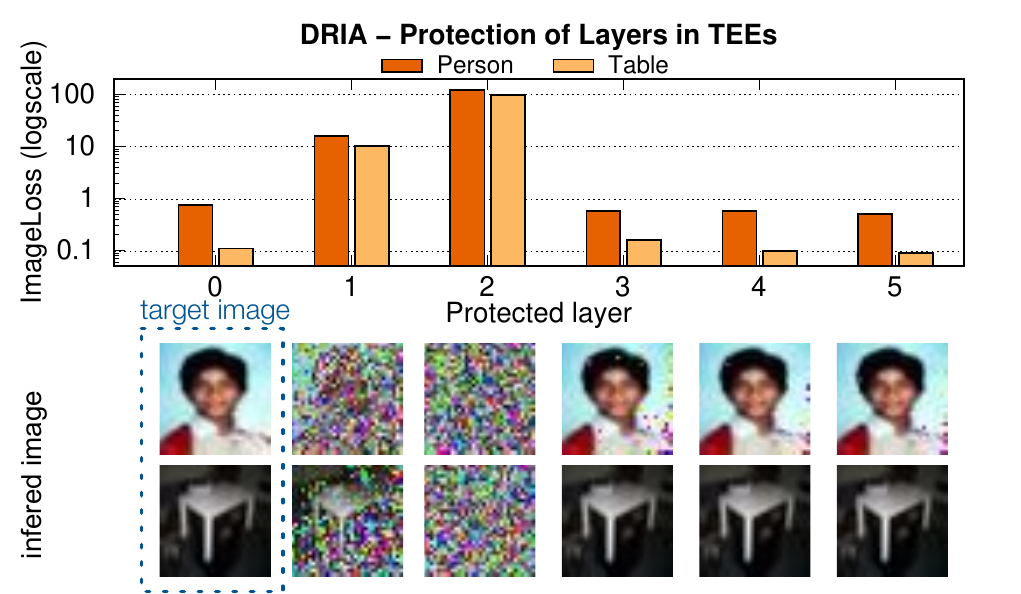}
        \caption{ImageLoss of two infered images on LeNet-5 using Static \sys}
    \end{subfigure}
    \begin{subfigure}{.5\textwidth}
        \vspace{10pt}
        \centering
        \includegraphics[scale=0.8]{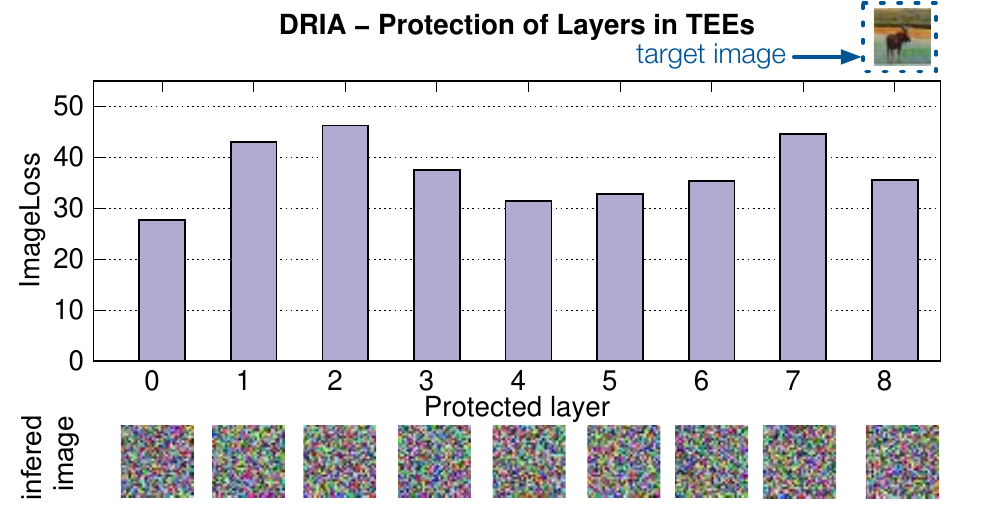}
        \caption{ImageLoss of an infered image on AlexNet using Static \sys}
    \end{subfigure}
    \caption{ImageLoss infered images with various protected layers using Static \sys.}
	%\vspace{-10pt}
    \label{fig:DRIA_results}
\end{figure}

\textbf{DRIA.} Securing the first layers (especially the $2{nd}$ layer) with static \sys is sufficient to make the attack fail in both our models. 
The attacker obtains a reconstructed yet blurry image with a large \textit{Image Loss} (see Figure~\ref{fig:DRIA_results}). 
Even if we were unable to reproduce a clear image with a non-protected AlexNet model, protecting layers inside a TEE enclave (in particular the second layer), still makes DRIA performs even worse. 
Indeed, for reconstructing visual data fed to a neural network, convolutional layers are the most suitable target for an attacker since they capture more of the visual features of data. 
Specifically, protecting the firsts convolutional layers (L1 + L2) have the highest impact against DRIA, since those are used to  extract the low-level visual features (\eg, image edges). 
Instead, the tail layers extract high-level features (\eg, color of eyes, shape complexity, \etc). 
By preventing the attacker from getting the low-level features, that are the support for the high level ones, it fails at rebuilding the input features.

\emph{Takeaway}: to mitigate DRIA, one should focus on securing the first layers of the convolutional part of the models.

%add figure for DRIA vs AlexNet 

\begin{figure}[!t]
    \begin{subfigure}{.5\textwidth}
        \centering
    	\includegraphics[scale=0.62,angle=270]{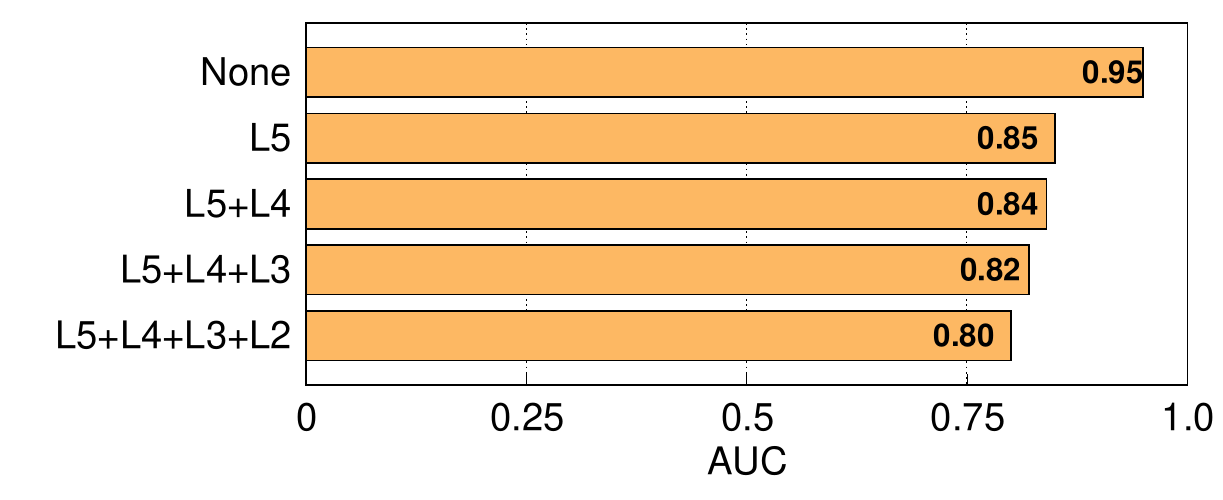}
    	\caption{AUC of MIA in LeNet-5 with various protected layers using Static \sys.}
    \end{subfigure}
    
    \begin{subfigure}{.5\textwidth}
        \centering
    	\includegraphics[scale=0.62,angle=270]{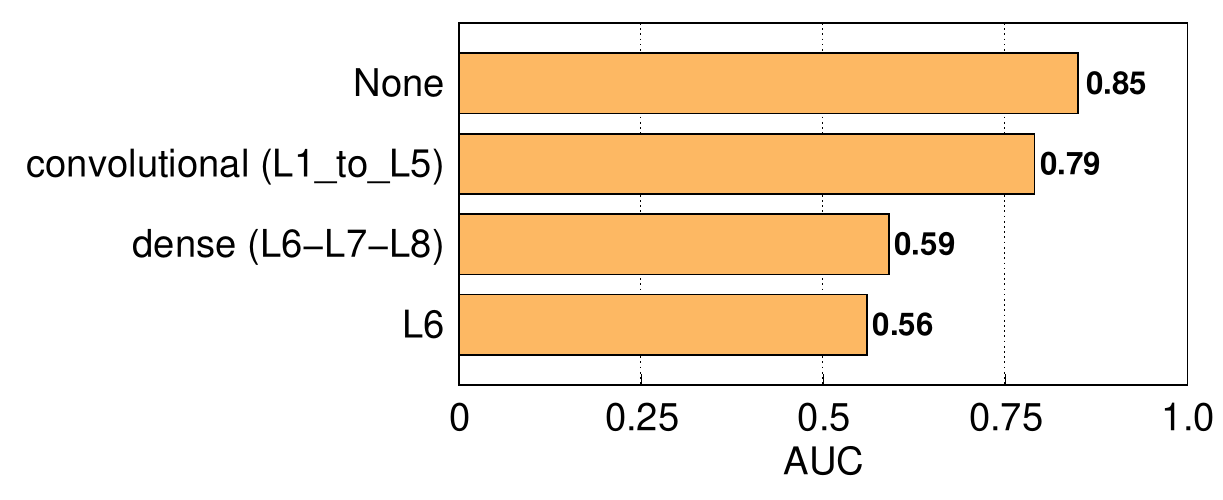}
    	\caption{AUC of MIA in AlexNet with various protected layers using Static \sys.}
    \end{subfigure}
    
    \caption{AUC of MIA with various protected layers using Static \sys.}
    \vspace{-5mm}
    \label{fig:MIA_results}
\end{figure}

\textbf{MIA.} Securing the last layer (\ie, the $5{th}$ layer) in LeNet-5 with static \sys lowers the attack's \textit{AUC} from 0.95 to 0.85. 
Protecting additional tail layers show little benefits, as the \textit{AUC} attack only drop by 5\% with last 4 layers protected (see Figure~\ref{fig:MIA_results} \textit{(a)}.
For AlexNet, the focus of an attacker should also be on the gradients of tail layers (L6, L7 and L8) that constitutes the dense part of the model, to succeed in MIA with an \textit{AUC} of 0.79. Thus, protecting these specific layers is effective to lower the \textit{AUC} until 0.59, assuming the attacker would exploit the gradients of the convolutional part of the model. 
To decrease the number of the protected layers, shielding layer L6 is sufficient to reduce the \textit{AUC} to 0.56 if the attacker focuses on exploiting the two remaining dense layers. 
Figure~\ref{fig:MIA_results} \textit{(b)} summarizes these results.

\emph{Takeaway}: to reduce MIA impact, securing layers of the Dense part usually found at the end of a model remains more efficient than securing the layers of the convolutional part.  

\begin{comment}
    \begin{figure}[!t]
        \centering
    	\includegraphics[scale=0.62,angle=270]{plots/MIA/mia.pdf}
        \vspace{-5pt}
        \caption{AUC of MIA in LeNet-5 with various protected layers using Static \sys.} 
        \vspace{-5mm}
        \label{fig:MIA_results}
    \end{figure}
    
    \begin{figure}[!t]
        \centering
    	\includegraphics[scale=0.62,angle=270]{plots/MIA_Alexnet/mia_alexnet.pdf}
        \vspace{-5pt}
        \caption{AUC of MIA in AlexNet with various protected layers using Static \sys.} 
        \vspace{-5mm}
        \label{fig:MIA_alexnet_results}
    \end{figure}
\end{comment}

\textbf{DPIA.} Protecting individual layers using \emph{static} \sys proves ineffective against this attacks. We systematically hit an AUC rate of 0.99 regardless the protected layer. 
While it is possible to lower the \textit{AUC} down to 0.85 with 4 protected layers inside the enclave, this would consumes a large share of the available secure memory (1.841MB for LeNet-5, see Section \ref{ssec:overhead}), a scarce  resource shared with other Secure Applications, and heavily impacts the training time (see~\ref{ssec:overhead}).
Instead, dynamic \sys achieves a better AUC rate (0.78) with only two simultaneous layers inside the enclave ($size_{MW}=2$) and with an appropriate choice of $V_{MW} ([0.2, 0.1, 0.6, 0.1])$.
Table~\ref{tab:dpia_stat} resumes these results.
To find the best distribution of $V_{MW}$ for each value of $size_{MW}$, we train different instances of the attack model (random forest) on a gradient train set with differently located missing data to reflect changing protected layers across the FL cycles, according to the chosen $V_{MW}$. 
The incomplete columns of the train set are filled with the mean strategy. 
We evaluate each attack model instance on a gradient validation set and we retain the $V_{MW}$ distribution of the worst instance. 
Finally, we test $V_{MW}$ on a gradient test set that reflects unseen gradients.

\begin{comment}
    \begin{table}[h!]
        \begin{center}
        \begin{tabular}{|c| c| c| c |c |c|} 
         \hline
         & \textbf{None} & \textbf{L4} & \textbf{L3+L4} & \textbf{L2+L3+L4} & \textbf{L1+L2+L3+L4}\\ [0.5ex]
         \hline
         \textbf{AUC} & 0.94 & 0.88 & 0.83 & 0.79 & 0.70\\ 
         \hline
        \end{tabular}
        \end{center}
        \caption{AUC of DPIA using Static \sys}
        \label{tab:dpia_stat}
        \vspace{-8mm}
    \end{table}
    
    \begin{table}[h!]
        \begin{center}
        \begin{tabular}{|c |c |c |c |c|} 
         \hline
         & \textbf{None} & \textbf{MW=2} & \textbf{MW=3} & \textbf{MW=4} \\ [0.5ex]
         \hline
         \textbf{AUC} & 0.94 & 0.71 & 0.64 & 0.62 \\ 
         \hline
        \end{tabular}
        \end{center}
        \caption{AUC of DPIA using Dynamic \sys}
        \label{tab:dpia_dyn}
        \vspace{-8mm}
    \end{table}
\end{comment}

\begin{table}[b!]
    \begin{center}
    \begin{tabular}{cp{0.75cm}p{0.5cm}p{0.9cm}cc} 
     
     \rowcolor{gray!50}
     \multicolumn{6}{|c|}{\textbf{Static \sys}} \\
     
     \rowcolor{gray!25}
     & \textbf{None} & \textbf{L4} & \textbf{L3+L4} & \textbf{L3+L4+L5} & \textbf{L2+L3+L4+L5}\\ [0.5ex]
     
     \textbf{AUC} & 0.99 & 0.99 & 0.99 & 0.95 & 0.85\\ 
     
     \rowcolor{gray!50}
     \multicolumn{6}{c}{\textbf{Dynamic \sys}} \\
     
     \rowcolor{gray!25}
     & \textbf{None} & \multicolumn{2}{c}{\textbf{MW=2}} & \textbf{MW=3} & \textbf{MW=4}\\ [0.5ex]
     
     \textbf{AUC} & 0.99 & \multicolumn{2}{c}{0.78} & 0.77 & 0.80\\
     
    \end{tabular}
    \end{center}
    \caption{AUC of DPIA using \sys}
    \label{tab:dpia_stat}
    \vspace{-8mm}
\end{table}

%\textbf{Grouped protection.}
%\vs{why this paragraph is here? Is this evaluated?} Given its ability to protect non-successive layers, \sys has another important advantage over its competitor: it is able to simultaneously minimize the impact of both DRIA and MIA, without the need to protect all intermediate layers as required in DarkneTZ.
       
%%%%%%%%%%%%%%%%%%%%%%%%%%%%%%%%%%%%%%%%%%%%%%%%%%%%%%%%%%%%%%%%%%%%%%%%%%%%
\begin{table*}[!t]
    \centering
    \rowcolors{1}{gray!10}{gray!0}
    \begin{tabular}{lp{5.8cm}p{2.2cm}}
         \rowcolor{gray!50}
         \textbf{Protected layers} & \textbf{CPU Training time (User+ Kernel+ Allocation)} & \textbf{TEE Memory Usage (at exec) in MB}\\
         
         \textbf{Without (Baseline)} & 2.191s + 0.021s + 0s & 0 \\
         
         \rowcolor{gray!30}
         \multicolumn{3}{c}{\textbf{Static \sys}}\\
         
         L1 & 1.886s + 0.738s + 0.09s \textbf{(19\% overhead)} & 1.127 \\
         
         \textbf{L2 (against DRIA)} & 1.672s + 0.652s + 0.34s \textbf{(20\% overhead)} & 0.565 \\
         
         L3 & 1.696s + 0.674s + 0.34s \textbf{(22\% overhead)} & 0.286 \\
         
         L4 & 1.691s + 0.673s + 0.34s \textbf{(22\% overhead)}& 0.286 \\
         
         \textbf{L5 (against MIA)} & 2.044s + 0.187s + 4.68s \textbf{(212\% overhead)} & 0.704 \\
         
         \textbf{L2+L5 (against DRIA+MIA)} & 1.561s + 0.846s + 5.02s \textbf{(235\% overhead)} & 1.269\\
         
         \rowcolor{gray!30}
         \multicolumn{3}{c}{\textbf{Dynamic \sys}}\\
         
         \rowcolor{gray!20}
         \multicolumn{3}{c}{\textbf{MW=2}}\\
         
         L1+L2 & 1.323s + 1.331s + 0.43s \textbf{(39\% overhead)} & 1.692\\
          
         L2+L3 & 1.139s + 1.275s + 0.68s \textbf{(40\%overhead)} & 0.851\\
          
         L3+L4 & 1.134s + 1.269s + 0.68s \textbf{(39\%overhead)} & 0.572\\
         
         L4+L5 & 1.507s + 0.808s + 5.02s \textbf{(231\%overhead)} & 0.99\\
         
         \textbf{AVG ($V_{MW}$ =[0.2, 0.1, 0.6, 0.1]) (against DPIA)}& 1.21s + 1.236s + 1.064s \textbf{(58.3\% overhead)} & 1.692 (AVG=0.866) \\
         
         \rowcolor{gray!20}
         \multicolumn{3}{c}{\textbf{MW=3}}\\
         
         L1+L2+L3 & 0.708s + 2.081s + 0.77s \textbf{(61\% overhead)} & 1.978\\
         
         L2+L3+L4 & 0.807s + 1.743s + 1.02s \textbf{(61\% overhead)} & 1.137\\
         
         L3+L4+L5 & 1.003s + 1.418s + 5.36s \textbf{(251\% overhead)} & 1.276\\
         
         \textbf{AVG ($V_{MW}$= [0.1, 0.1, 0.8]) (against DPIA)} & 0.964s + 1.517s + 4.467s \textbf{(213\% overhead)} & 1.978 (AVG=1.332)\\
         
         \rowcolor{gray!20}
         \multicolumn{3}{c}{\textbf{MW=4}}\\
         
         L1+L2+L3+L4 & 0.170s + 2.754s + 1.11s \textbf{(82\% overhead)} & 2.264\\
         
         L2+L3+L4+L5 & 0,985s + 1,420s + 5.7s \textbf{(266\% overhead)} & 1.841\\
         
         \textbf{AVG ($V_{MW}$ =[0.1 0.9])} & 0.904s + 1.553s + 5.241s \textbf{(247\% overhead)} & 2.264 (AVG=1.883)\\
    \end{tabular}
    \caption{CPU Time and TEE memory usage of \sys (LeNet-5, CIFAR-100, batch-size = 32)}
    \label{tab:Performance tests}
\end{table*}
	   
\subsection{Overhead}\label{ssec:overhead}
We use two metrics to measure the performance impact of \sys on the LeNet-5 model. 
Firstly, we consider the model training time of one FL cycle per configuration of protected layer(s). 
We break down this in three parts: \emph{(1)} computation in user-space, spent outside the TrustZone enclave, \emph{(2)} kernel-space time, spent inside the enclave during the training process, and \emph{(3)} allocation time, \ie time to allocate the TEE memory for the received weights, before starting the training process. 
We rely on the real-time dashboard provided by DarkneTZ to measure the user and kernel time. 
For the TEE memory allocation time, we instrumented the code with timers around the memory allocation for the model weights.
 
\begin{figure*}[!t]
    \centering
	\includegraphics[scale=0.4]{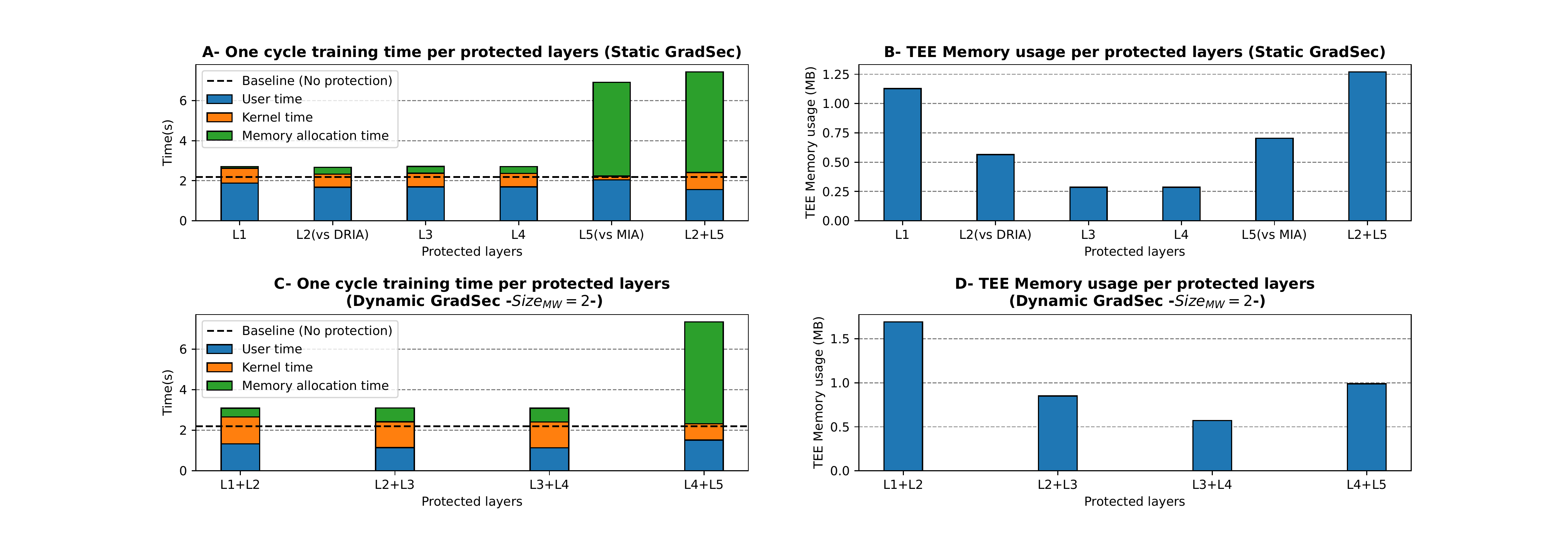}
    \caption{Training time (A,C) and TEE memory usage (B, C) for static and dynamic \sys for different numbers of protected layers.} 
    %\vspace{-5mm}
    \label{fig:Performance tests}
\end{figure*}

Secondly, we measure the maximum TEE memory usage for each configuration of protected layer(s).
TrustZone doesn't provide a direct tool to measure the used TEE memory size and doing so from the \textit{normal world} is prohibited due to the restriction imposed by the \textit{secure world}. 
To address this issue, we locate all the DarkneTZ code parts where the TEE memory allocation is triggered through the \textit{malloc} / \textit{calloc} instructions and summed the sizes of allocated regions.
%we studied the inner architecture of each protected layer to determine its weights size. We figure out further the DNN layer components that significantly influence the established measures.
In the case of dynamic \sys, we computed the previous metrics for each $size_{MW}$ and using the best $V_{MW}$ distribution (\ie, the one achieving the highest security). 
Since the protected layers change over the FL cycles, we use a weighted average over all the different protection positions covered by the \textit{MW} to compute the training time. 
For the TEE memory usage, we just included the consumption of the most expensive combination among the different possibilities allowed by the \textit{MW}. 
The results are summarized in the Table~\ref{tab:Performance tests} and Figure~\ref{fig:Performance tests}, and discussed in the remainder of this section.

\textbf{DRIA.}
To protect LeNet-5 against DRIA, static \sys should focus on securing the layer L2. 
Weobserve a 20\% increase in the training time when compared to training the model outside the enclave, while the required TEE memory size is about 0.57MB.

\textbf{MIA.}
To protect LeNet-5 against MIA, static \sys should focus on securing the last layer (L5). 
By doing so, we measure an important increase in the training time (\ie 212\%) when compared to training the model outside the enclave, while the required TEE memory size is about 0.70MB. 
This significant overhead for the training time is mainly caused by the memory allocation for L5 which has a fairly large number of parameters (76.8K). 

\textbf{Grouped protection.}
Protecting LeNet-5 against DRIA and MIA simultaneously require to secure the sensitive layer of each attack with Static \sys, \ie L2 and L5. 
Unsurprisingly, the training time overhead increases by 235\%, while the required TEE memory is 1.27MB, the fairly heavy L5 being responsible for the large majority of this overhead. 
%Using DarkneTZ for this grouped protection require us to secure many layers starting from L2 to L5. With the cost of 4 layers inside the enclave, we reach 266\% overhead. 

\textbf{DPIA.}
To mitigate DPIA with dynamic \sys, the best option in terms of security and overhead is to use $size_{MW}=2$ with $V_{MW}=[0.2, 0.1, 0.6, 0.1]$. Indeed, as we have seen in section \ref{ssec:security_analysis}, this configuration offers better protection than DarkneTZ while requiring only two simultaneously protected layers for each cycle, 
The overhead to mitigate DPIA is a 53\% longer training time and 1.692 MB of TEE memory usage in the worst case (when the $MW$ protects L1+L2).

\begin{figure*}[h]
    \centering
	\includegraphics[scale=0.4,trim=80pt 0 0 0]{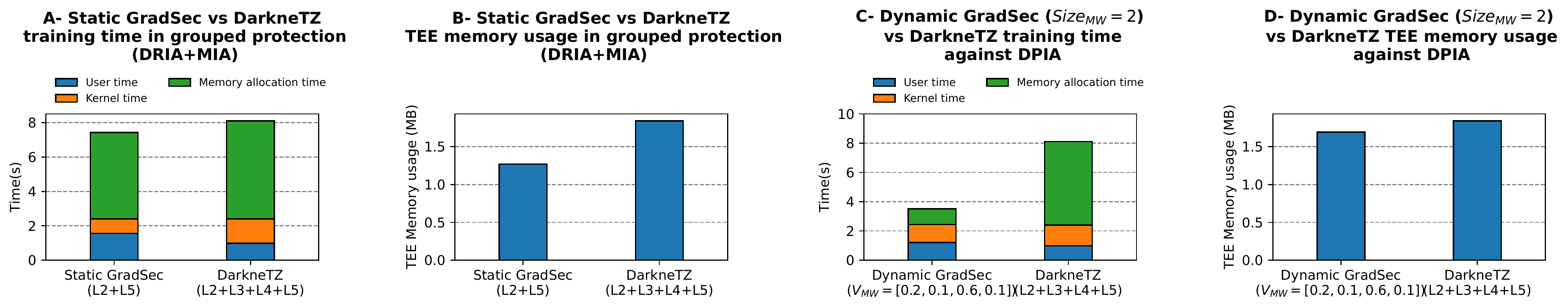}
    \caption{Comparison of training time and memory usage between static (A,B), dynamic (C,D) \sys and DarkneTZ.} 
    %\vspace{-5mm}
    \label{fig:XYZvsDarkneTZ}
\end{figure*}

\textbf{Comparison with DarkneTZ.}
Finally, we compare \sys against DarkneTZ in terms of training time and TEE memory usage.
The results are shown Figure~\ref{fig:XYZvsDarkneTZ}. 
We use static \sys to offer simultaneous protection against DRIA and MIA with the cost of 2 protected layers (L2 and L5). 
To offer a comparable level of protection with DarkneTZ, not only L2 and L5 but also all the intermediate layers (L3 and L4) should be protected in secured memory.
%This additional cost is due to the technical limitation of DarkneTZ which cannot protect non-successive layers. 
As one can imagine, static \sys offers a better training time (8.3\%) and significantly less TEE memory usage (30\%), due to the reduced numbers of layers to protect.
Concerning DPIA, we compare dynamic \sys with $size_{MW}=2$ and the most appropriate $V_{MW}$ (\ie $[0.2, 0.1, 0.6, 0.1]$), against DarkneTZ with the layers L2 up to L5 in the TEE enclave. Our flexible and more customizable approach allows us to find suitable $size_{MW}$ and $V_{MW}$ that deliver better level of protection than DarkneTZ and with less overhead. 
Indeed, changing the protected layers dynamically allows us to reduce the training time by ~56\% compared to DarkenTZ, mainly because \sys  does not require to allocate the necessary TEE memory for the biggest layer (L5) in each FL cycle. 
The necessary TEE memory for dynamic \sys varies according to the layers protected by the Moving Window in the current cycle. 
The most expensive configuration is when the Moving Window secures L1+L2. 
However, even in this configuration, dynamic \sys consumes 8\% less memory than DarkneTZ.

%*****************************************************************************

%% file: relwork.tex
\section{Related Work}\label{sec:rw}
We survey related work in the area of secure, confidential or privacy-preserving federated learning systems.

\noindent \textbf{PPFL.}
Fan Mo's PPFL (Privacy-preserving Federated Learning)~\cite{mo2022ppfl} is a TEE-based framework for mobile systems that limits privacy leakage in FL. 
Similar to \sys, PPFL aims at hiding the gradient updates of the model inside a TEE enclave on the client side for local training as well as on the server side for secure aggregation.
It leverages layer-wise training to train each model’s layer separately inside the trusted area until its convergence. 
The PPFL protocol implies the modification of the FL process to support layer-wise distribution of the model to the clients instead of distributing the whole model at once. 
It also implies the constant use of TEEs to train each layer of the model. 
While this solution is sound from a privacy perspective, it also incurs a substantial overhead in terms of training time by design as model layers are trained in a sequential manner.

\noindent \textbf{Gecko.}
Gecko training~\cite{duddu2022towards} is a methodology developed to bring privacy-aware deep learning for embedded systems. 
Its goal is to offer membership-privacy by design in neural networks by leveraging quantization to reconcile privacy, accuracy and efficiency. 
The main objective of the Gecko design is to mitigate blackbox MIA. 
However, contrary to \sys, no evidence is given regarding the resilience of the proposed solution against other attacks we dealt with (\eg, DRIA, DPIA).

\noindent \textbf{BatchCrypt}~\cite{zhang2020batchcrypt} is an Homomorphic Encryption (HE) method for Cross-Silo FL. 
It aims at reducing the cost in computation and communication of HE when the gradients are homorphically encrypted. 
Instead of encrypting individual gradients with full precision, BatchCrypt encodes a batch of quantized gradients into a long integer and encrypt it in one go. 
To allow aggregation of gradients to be performed on ciphertexts of the encoded batches, authors in~\cite{zhang2020batchcrypt} propose new quantization and encoding schemes, alongside a new gradient clipping technique. 
BatchCrypt considerably improves over vanilla HE method and is well suited to ML constraints.
However, it lacks support for an end-to-end solution to circumvent the problem of comprised clients whose OS may leak the gradients before being encrypted, in contrast with TEE solutions like \sys. 

\noindent \textbf{Slalom}~\cite{tramer2018slalom} presents a system and solution for high performance execution of Deep Neural Networks (DNNs) in TEEs.
In a nutshell, it efficiently partitions DNN computations between trusted and untrusted devices. 
It leverages both GPU (for efficient batch computation) and a TEE (for minimizing the use of cryptography). 
Specifically, Slalom is a framework that delegates execution of all dense layers in a DNN from a TEE to a faster, yet untrusted, processor (\ie, typically, a GPU). 
It requires a lot of pre-computation over known and fixed weights, and hence it only supports private inference and not training. 
In addition, dense layers may also leak sensitive information, useful for MIA, especially if the computation in which they are involved are computed outside the TEE enclave.

\noindent \textbf{Citadel}~\cite{zhang2021citadel} is a federated learning framework, built on top of Intel SGX enclaves.
It relies on two distinguished set of worker enclaves (\emph{training} and \emph{aggregator} ones).
It employs zero-sum masking and hierarchical aggregation techniques to prevent gradient leaking across such worker enclaves.
However, it does not protect against membership-inference attacks, and it is strongly coupled to Intel SGX enclaves. 
Given the future roadmap of Intel toward server-only deployments of SGX, we believe \sys to be better suited to be deployed on edge devices in a federated learning system.
%\section*{Acknowledgements}
%This work was partly supported by the European Commission through the H2020 VEDLIoT project (957197).

%% file: conclusion.tex
\section{Conclusion and Future Work}\label{sec:conclusion}
We presented \sys, a TEE-based protection mechanism that improves FL privacy guarantees against state-of-the-art inference attacks.
\sys can operate in two modes: static and dynamic.
Static \sys can simultaneously protect against DRIA and MIA attacks with less overhead than DarkneTZ.
Dynamic \sys offers a better protection than DarkneTZ against DPIA while still incurring less overhead.
We implemented \sys on top of the OP-TEE trusted OS, and evaluated its performance on a ARM Cortex-A53 processor with support for TrustZone enclaves.
We plan to release \sys to the research and open-source community.

We intend to extend this work along the following directions.
First, we aim at adding support for RNNs (Recurrent neural networks). This would allow us to protect other types of machine learning models (\eg, models dealing with text, voice recordings and time series in general), and validate our approach on highly-sensitive domains (\eg, e-health).
Second, we intend to study hybrid deployments, in which TEE-enabled clients are deployed alongside legacy clients without support for TEEs, and for which purely software-based approaches are necessary.

%\section*{Acknowledgements}
%This work was partly supported by the European Commission through the H2020 VEDLIoT project (957197).

%% file: main_file.bbl

\begin{thebibliography}{59}

%%% the Web address.

\ifx \showCODEN    \undefined \def \showCODEN     #1{\unskip}     \fi
\ifx \showDOI      \undefined \def \showDOI       #1{#1}\fi
\ifx \showISBNx    \undefined \def \showISBNx     #1{\unskip}     \fi
\ifx \showISBNxiii \undefined \def \showISBNxiii  #1{\unskip}     \fi
\ifx \showISSN     \undefined \def \showISSN      #1{\unskip}     \fi
\ifx \showLCCN     \undefined \def \showLCCN      #1{\unskip}     \fi
\ifx \shownote     \undefined \def \shownote      #1{#1}          \fi
\ifx \showarticletitle \undefined \def \showarticletitle #1{#1}   \fi
\ifx \showURL      \undefined \def \showURL       {\relax}        \fi
% invisible to TeX
\providecommand\bibfield[2]{#2}
\providecommand\bibinfo[2]{#2}
\providecommand\natexlab[1]{#1}
\providecommand\showeprint[2][]{arXiv:#2}

\bibitem[Albawi et~al\mbox{.}(2017)]%
        {albawi2017understanding}
\bibfield{author}{\bibinfo{person}{Saad Albawi}, \bibinfo{person}{Tareq~Abed
  Mohammed}, {and} \bibinfo{person}{Saad Al-Zawi}.}
  \bibinfo{year}{2017}\natexlab{}.
\newblock \showarticletitle{Understanding of a convolutional neural network}.
  In \bibinfo{booktitle}{\emph{2017 International Conference on Engineering and
  Technology (ICET)}} (Nashville, TN, USA, August 21-24).
  \bibinfo{publisher}{IEEE}, \bibinfo{address}{Manhattan, NY, USA},
  \bibinfo{pages}{1--6}.
\newblock
\urldef\tempurl%
\url{https://doi.org/10.1109/ICEngTechnol.2017.8308186}
\showDOI{\tempurl}


\bibitem[Amacher and Schiavoni(2019)]%
        {amacher2019performance}
\bibfield{author}{\bibinfo{person}{Julien Amacher} {and}
  \bibinfo{person}{Valerio Schiavoni}.} \bibinfo{year}{2019}\natexlab{}.
\newblock \showarticletitle{On the Performance of {ARM} TrustZone - (Practical
  Experience Report)}. In \bibinfo{booktitle}{\emph{Distributed Applications
  and Interoperable Systems - 19th {IFIP} {WG} 6.1 International Conference,
  {DAIS} 2019, Held as Part of the 14th International Federated Conference on
  Distributed Computing Techniques, DisCoTec 2019, Kongens Lyngby, Denmark,
  Proceedings}} (Kongens Lyngby, Denmark, June 17-21)
  \emph{(\bibinfo{series}{Lecture Notes in Computer Science})}.
  \bibinfo{publisher}{Springer}, \bibinfo{address}{New York, NY, USA},
  \bibinfo{pages}{133--151}.
\newblock
\urldef\tempurl%
\url{https://doi.org/10.1007/978-3-030-22496-7\_9}
\showDOI{\tempurl}


\bibitem[Android(2020)]%
        {web:trusty}
\bibfield{author}{\bibinfo{person}{Android}.} \bibinfo{year}{2020}\natexlab{}.
\newblock \bibinfo{booktitle}{\emph{Trusty TEE}}.
\newblock
\urldef\tempurl%
\url{https://source.android.com/security/trusty?hl=en}
\showURL{%
Retrieved October 8, 2022 from \tempurl}


\bibitem[ARM(2022)]%
        {web:armwebsite}
\bibfield{author}{\bibinfo{person}{ARM}.} \bibinfo{year}{2022}\natexlab{}.
\newblock \bibinfo{booktitle}{\emph{Arm CPU Architecture: A Foundation for
  Computing Everywhere}}.
\newblock
\urldef\tempurl%
\url{https://www.arm.com/architecture/cpu}
\showURL{%
Retrieved October 8, 2022 from \tempurl}


\bibitem[Benedito et~al\mbox{.}(2021)]%
        {benedito2021kevlar}
\bibfield{author}{\bibinfo{person}{Oscar Benedito}, \bibinfo{person}{Ricard
  Delgado{-}Gonzalo}, {and} \bibinfo{person}{Valerio Schiavoni}.}
  \bibinfo{year}{2021}\natexlab{}.
\newblock \showarticletitle{KeVlar-Tz: {A} Secure Cache for ArmTrustZone -
  (Practical Experience Report)}. In \bibinfo{booktitle}{\emph{Distributed
  Applications and Interoperable Systems - 21st {IFIP} {WG} 6.1 International
  Conference, {DAIS} 2021, Held as Part of the 16th International Federated
  Conference on Distributed Computing Techniques, DisCoTec 2021 Proceedings}}
  (Valletta, Malta, June 14-18), \bibfield{editor}{\bibinfo{person}{Miguel
  Matos} {and} \bibinfo{person}{Fab{\'{\i}}ola Greve}} (Eds.).
  \bibinfo{publisher}{Springer}, \bibinfo{address}{New York, NY, USA},
  \bibinfo{pages}{109--124}.
\newblock
\urldef\tempurl%
\url{https://doi.org/10.1007/978-3-030-78198-9\_8}
\showDOI{\tempurl}


\bibitem[Bharath~Ramsundar(2022)]%
        {web:denselayer}
\bibfield{author}{\bibinfo{person}{Reza Bosagh~Zadeh Bharath~Ramsundar}.}
  \bibinfo{year}{2022}\natexlab{}.
\newblock \bibinfo{booktitle}{\emph{Fully Connected Deep Networks}}.
\newblock
\urldef\tempurl%
\url{https://www.oreilly.com/library/view/tensorflow-for-deep/9781491980446/ch04.html}
\showURL{%
Retrieved October 8, 2022 from \tempurl}


\bibitem[Bhargav-Spantzel(2014)]%
        {bhargav2014trusted}
\bibfield{author}{\bibinfo{person}{Abhilasha Bhargav-Spantzel}.}
  \bibinfo{year}{2014}\natexlab{}.
\newblock \showarticletitle{Trusted Execution Environment for Privacy
  Preserving Biometric Authentication}.
\newblock \bibinfo{journal}{\emph{Intel Technology Journal}}
  \bibinfo{volume}{18}, \bibinfo{number}{4} (\bibinfo{year}{2014}),
  \bibinfo{numpages}{16}~pages.
\newblock


\bibitem[Bonawitz et~al\mbox{.}(2017)]%
        {bonawitz2017practical}
\bibfield{author}{\bibinfo{person}{Keith Bonawitz}, \bibinfo{person}{Vladimir
  Ivanov}, \bibinfo{person}{Ben Kreuter}, \bibinfo{person}{Antonio Marcedone},
  \bibinfo{person}{H.~Brendan McMahan}, \bibinfo{person}{Sarvar Patel},
  \bibinfo{person}{Daniel Ramage}, \bibinfo{person}{Aaron Segal}, {and}
  \bibinfo{person}{Karn Seth}.} \bibinfo{year}{2017}\natexlab{}.
\newblock \showarticletitle{Practical Secure Aggregation for Privacy-Preserving
  Machine Learning}. In \bibinfo{booktitle}{\emph{Proceedings of the 2017 ACM
  SIGSAC Conference on Computer and Communications Security}} (Dallas, TX, USA,
  October 30 - November 3). \bibinfo{publisher}{ACM}, \bibinfo{address}{New
  York, NY, USA}, \bibinfo{pages}{1175–1191}.
\newblock
\urldef\tempurl%
\url{https://doi.org/10.1145/3133956.3133982}
\showDOI{\tempurl}


\bibitem[Bonawitz et~al\mbox{.}(2019)]%
        {bonawitz2019towards}
\bibfield{author}{\bibinfo{person}{Kallista~A. Bonawitz},
  \bibinfo{person}{Hubert Eichner}, \bibinfo{person}{Wolfgang Grieskamp},
  \bibinfo{person}{Dzmitry Huba}, \bibinfo{person}{Alex Ingerman},
  \bibinfo{person}{Vladimir Ivanov}, \bibinfo{person}{Chlo{\'{e}} Kiddon},
  \bibinfo{person}{Jakub Kone{\v{c}}n{\'y}}, \bibinfo{person}{Stefano
  Mazzocchi}, \bibinfo{person}{H.~Brendan McMahan}, \bibinfo{person}{Timon~Van
  Overveldt}, \bibinfo{person}{David Petrou}, \bibinfo{person}{Daniel Ramage},
  {and} \bibinfo{person}{Jason Roselander}.} \bibinfo{year}{2019}\natexlab{}.
\newblock \showarticletitle{Towards Federated Learning at Scale: System
  Design}.
\newblock  (\bibinfo{year}{2019}).
\newblock
\urldef\tempurl%
\url{https://doi.org/10.48550/arXiv.1902.01046}
\showDOI{\tempurl}
\showeprint[arXiv]{1902.01046}


\bibitem[Bonawitz et~al\mbox{.}(2016)]%
        {bonawitz2016practical}
\bibfield{author}{\bibinfo{person}{Kallista~A. Bonawitz},
  \bibinfo{person}{Vladimir Ivanov}, \bibinfo{person}{Ben Kreuter},
  \bibinfo{person}{Antonio Marcedone}, \bibinfo{person}{H.~Brendan McMahan},
  \bibinfo{person}{Sarvar Patel}, \bibinfo{person}{Daniel Ramage},
  \bibinfo{person}{Aaron Segal}, {and} \bibinfo{person}{Karn Seth}.}
  \bibinfo{year}{2016}\natexlab{}.
\newblock \showarticletitle{Practical Secure Aggregation for Federated Learning
  on User-Held Data}.
\newblock  (\bibinfo{year}{2016}).
\newblock
\urldef\tempurl%
\url{https://doi.org/10.48550/arXiv.1611.04482}
\showDOI{\tempurl}
\showeprint[arXiv]{1611.04482}


\bibitem[Bradshaw(2021)]%
        {web:armincrease}
\bibfield{author}{\bibinfo{person}{Kyle Bradshaw}.}
  \bibinfo{year}{2021}\natexlab{}.
\newblock \bibinfo{booktitle}{\emph{Android now powers over 3 billion
  devices}}.
\newblock Google.
\newblock
\urldef\tempurl%
\url{https://9to5google.com/2021/05/18/android-now-powers-over-3-billion-devices/}
\showURL{%
Retrieved October 8, 2022 from \tempurl}


\bibitem[Brownlee(2019)]%
        {web:convolutionallayer}
\bibfield{author}{\bibinfo{person}{Jason Brownlee}.}
  \bibinfo{year}{2019}\natexlab{}.
\newblock \bibinfo{booktitle}{\emph{How Do Convolutional Layers Work in Deep
  Learning Neural Networks?}}
\newblock
\urldef\tempurl%
\url{https://machinelearningmastery.com/convolutional-layers-for-deep-learning-neural-networks/}
\showURL{%
Retrieved October 8, 2022 from \tempurl}


\bibitem[Costan and Devadas(2016)]%
        {costan2016intel}
\bibfield{author}{\bibinfo{person}{Victor Costan} {and}
  \bibinfo{person}{Srinivas Devadas}.} \bibinfo{year}{2016}\natexlab{}.
\newblock \showarticletitle{Intel sgx explained.}
\newblock \bibinfo{journal}{\emph{IACR Cryptol. ePrint Arch.}}
  \bibinfo{volume}{2016}, \bibinfo{number}{86} (\bibinfo{year}{2016}),
  \bibinfo{pages}{1--118}.
\newblock


\bibitem[Dilmegani(2022)]%
        {web:whatisfl}
\bibfield{author}{\bibinfo{person}{Cem Dilmegani}.}
  \bibinfo{year}{2022}\natexlab{}.
\newblock \bibinfo{booktitle}{\emph{What is Federated Learning (FL)? Techniques
  \& Benefits in 2022}}.
\newblock
\urldef\tempurl%
\url{https://research.aimultiple.com/federated-learning/}
\showURL{%
Retrieved October 8, 2022 from \tempurl}


\bibitem[Duddu et~al\mbox{.}(2022)]%
        {duddu2022towards}
\bibfield{author}{\bibinfo{person}{Vasisht Duddu}, \bibinfo{person}{Antoine
  Boutet}, {and} \bibinfo{person}{Virat Shejwalkar}.}
  \bibinfo{year}{2022}\natexlab{}.
\newblock \showarticletitle{Towards privacy aware deep learning for embedded
  systems}. In \bibinfo{booktitle}{\emph{{SAC} '22: The 37th {ACM/SIGAPP}
  Symposium on Applied Computing, Virtual Event}} (Virtual Event, April 25 -
  29), \bibfield{editor}{\bibinfo{person}{Jiman Hong},
  \bibinfo{person}{Miroslav Bures}, \bibinfo{person}{Juw~Won Park}, {and}
  \bibinfo{person}{Tom{\'{a}}s Cern{\'{y}}}} (Eds.). \bibinfo{publisher}{ACM},
  \bibinfo{address}{New York, NY, USA}, \bibinfo{pages}{520--529}.
\newblock
\urldef\tempurl%
\url{https://doi.org/10.1145/3477314.3507128}
\showDOI{\tempurl}


\bibitem[Dwork(2008)]%
        {dwork2008differential}
\bibfield{author}{\bibinfo{person}{Cynthia Dwork}.}
  \bibinfo{year}{2008}\natexlab{}.
\newblock \showarticletitle{Differential Privacy: {A} Survey of Results}. In
  \bibinfo{booktitle}{\emph{Theory and Applications of Models of Computation,
  5th International Conference, {TAMC} 2008}} (Xi'an, China, April 25-29)
  \emph{(\bibinfo{series}{Lecture Notes in Computer Science},
  Vol.~\bibinfo{volume}{4978})}, \bibfield{editor}{\bibinfo{person}{Manindra
  Agrawal}, \bibinfo{person}{Ding{-}Zhu Du}, \bibinfo{person}{Zhenhua Duan},
  {and} \bibinfo{person}{Angsheng Li}} (Eds.). \bibinfo{publisher}{Springer},
  \bibinfo{address}{New York, NY, USA}, \bibinfo{pages}{1--19}.
\newblock
\urldef\tempurl%
\url{https://doi.org/10.1007/978-3-540-79228-4\_1}
\showDOI{\tempurl}


\bibitem[Fang and Qian(2021)]%
        {fang2021privacy}
\bibfield{author}{\bibinfo{person}{Haokun Fang} {and} \bibinfo{person}{Quan
  Qian}.} \bibinfo{year}{2021}\natexlab{}.
\newblock \showarticletitle{Privacy preserving machine learning with
  homomorphic encryption and federated learning}.
\newblock \bibinfo{journal}{\emph{Future Internet}} \bibinfo{volume}{13},
  \bibinfo{number}{4} (\bibinfo{year}{2021}), \bibinfo{pages}{94}.
\newblock


\bibitem[Fine(2006)]%
        {fine2006feedforward}
\bibfield{author}{\bibinfo{person}{Terrence~L Fine}.}
  \bibinfo{year}{2006}\natexlab{}.
\newblock \bibinfo{booktitle}{\emph{Feedforward neural network methodology}}.
\newblock \bibinfo{publisher}{Springer Science \& Business Media},
  \bibinfo{address}{New York, NY}.
\newblock


\bibitem[Google(2021)]%
        {web:growingFL}
\bibfield{author}{\bibinfo{person}{Google}.} \bibinfo{year}{2021}\natexlab{}.
\newblock \bibinfo{booktitle}{\emph{Why Is Federated Learning Getting So
  Popular}}.
\newblock
\urldef\tempurl%
\url{https://trends.google.com/trends/explore?date=today\%205-y&q=federated\%20learning}
\showURL{%
Retrieved October 8, 2022 from \tempurl}


\bibitem[Hein et~al\mbox{.}(2015)]%
        {hein2015secure}
\bibfield{author}{\bibinfo{person}{Daniel~M. Hein}, \bibinfo{person}{Johannes
  Winter}, {and} \bibinfo{person}{Andreas Fitzek}.}
  \bibinfo{year}{2015}\natexlab{}.
\newblock \showarticletitle{Secure Block Device - Secure, Flexible, and
  Efficient Data Storage for {ARM} TrustZone Systems}. In
  \bibinfo{booktitle}{\emph{2015 {IEEE} TrustCom/BigDataSE/ISPA}} (Helsinki,
  Finland, August 20-22). \bibinfo{publisher}{IEEE},
  \bibinfo{address}{Manhattan, NY, USA}, \bibinfo{pages}{222--229}.
\newblock
\urldef\tempurl%
\url{https://doi.org/10.1109/Trustcom.2015.378}
\showDOI{\tempurl}


\bibitem[Hongyan~Chang(2019)]%
        {github:MIA_github}
\bibfield{author}{\bibinfo{person}{Reza~Shokri Hongyan~Chang, Martin~Strobel}.}
  \bibinfo{year}{2019}\natexlab{}.
\newblock \bibinfo{booktitle}{\emph{ML Privacy Meter}}.
\newblock github.
\newblock
\urldef\tempurl%
\url{https://github.com/privacytrustlab/ml_privacy_meter}
\showURL{%
Retrieved October 8, 2022 from \tempurl}


\bibitem[Huang et~al\mbox{.}(2007)]%
        {LFWTech}
\bibfield{author}{\bibinfo{person}{Gary~B. Huang}, \bibinfo{person}{Manu
  Ramesh}, \bibinfo{person}{Tamara Berg}, {and} \bibinfo{person}{Erik
  Learned-Miller}.} \bibinfo{year}{2007}\natexlab{}.
\newblock \bibinfo{booktitle}{\emph{Labeled Faces in the Wild: A Database for
  Studying Face Recognition in Unconstrained Environments}}.
\newblock \bibinfo{type}{{T}echnical {R}eport} 07-49.
  \bibinfo{institution}{University of Massachusetts, Amherst}.
\newblock


\bibitem[Jang et~al\mbox{.}(2015)]%
        {jang2015secret}
\bibfield{author}{\bibinfo{person}{Jin~Soo Jang}, \bibinfo{person}{Sunjune
  Kong}, \bibinfo{person}{Minsu Kim}, \bibinfo{person}{Daegyeong Kim}, {and}
  \bibinfo{person}{Brent~ByungHoon Kang}.} \bibinfo{year}{2015}\natexlab{}.
\newblock \showarticletitle{SeCReT: Secure Channel between Rich Execution
  Environment and Trusted Execution Environment}. In
  \bibinfo{booktitle}{\emph{22nd Annual Network and Distributed System Security
  Symposium, {NDSS} 2015, San Diego, California, USA, February 8-11, 2015}}
  (San Diego, California, USA). \bibinfo{publisher}{The Internet Society},
  \bibinfo{address}{Reston, Virginia, USA}, \bibinfo{numpages}{15}~pages.
\newblock
\urldef\tempurl%
\url{https://www.ndss-symposium.org/ndss2015/secret-secure-channel-between-rich-execution-environment-and-trusted-execution-environment}
\showURL{%
\tempurl}


\bibitem[Kanagavelu et~al\mbox{.}(2021)]%
        {kanagavelu2021federated}
\bibfield{author}{\bibinfo{person}{Renuga Kanagavelu},
  \bibinfo{person}{Zengxiang Li}, \bibinfo{person}{Juniarto Samsudin},
  \bibinfo{person}{Shaista Hussain}, \bibinfo{person}{Feng Yang},
  \bibinfo{person}{Yechao Yang}, \bibinfo{person}{Rick Siow~Mong Goh}, {and}
  \bibinfo{person}{Mervyn Cheah}.} \bibinfo{year}{2021}\natexlab{}.
\newblock \bibinfo{booktitle}{\emph{Federated Learning for Advanced
  Manufacturing Based on Industrial IoT Data Analytics}}.
\newblock \bibinfo{publisher}{Springer International Publishing},
  \bibinfo{address}{Cham}, \bibinfo{pages}{143--176}.
\newblock
\showISBNx{978-3-030-67270-6}
\urldef\tempurl%
\url{https://doi.org/10.1007/978-3-030-67270-6_6}
\showDOI{\tempurl}


\bibitem[Kaplan et~al\mbox{.}(2016)]%
        {whitepaper:kaplan2016amd}
\bibfield{author}{\bibinfo{person}{David Kaplan}, \bibinfo{person}{Jeremy
  Powell}, {and} \bibinfo{person}{Tom Woller}.}
  \bibinfo{year}{2016}\natexlab{}.
\newblock \bibinfo{booktitle}{\emph{AMD memory encryption}}.
\newblock AMD.
\newblock
\urldef\tempurl%
\url{https://developer.amd.com/wordpress/media/2013/12/AMD_Memory_Encryption_Whitepaper_v9-Public.pdf}
\showURL{%
Retrieved October 8, 2022 from \tempurl}


\bibitem[Kingma and Ba(2015)]%
        {kingma2014adam}
\bibfield{author}{\bibinfo{person}{Diederik~P Kingma} {and}
  \bibinfo{person}{Jimmy Ba}.} \bibinfo{year}{2015}\natexlab{}.
\newblock \showarticletitle{Adam: A method for stochastic optimization}. In
  \bibinfo{booktitle}{\emph{3rd International Conference on Learning
  Representations, {ICLR} 2015, Conference Track Proceedings}} (San Diego, CA,
  USA, May 7-9). \bibinfo{publisher}{arXiv.org}, \bibinfo{address}{Ithaca, NY,
  USA}, \bibinfo{numpages}{13}~pages.
\newblock


\bibitem[Kostiainen et~al\mbox{.}(2011)]%
        {kostiainen2011practical}
\bibfield{author}{\bibinfo{person}{Kari Kostiainen}, \bibinfo{person}{N.
  Asokan}, {and} \bibinfo{person}{Jan{-}Erik Ekberg}.}
  \bibinfo{year}{2011}\natexlab{}.
\newblock \showarticletitle{Practical Property-Based Attestation on Mobile
  Devices}. In \bibinfo{booktitle}{\emph{Trust and Trustworthy Computing - 4th
  International Conference, {TRUST}}} (Pittsburgh, PA, USA, June 22-24).
  \bibinfo{publisher}{Springer}, \bibinfo{address}{New York, NY, USA},
  \bibinfo{pages}{78--92}.
\newblock
\urldef\tempurl%
\url{https://doi.org/10.1007/978-3-642-21599-5\_6}
\showDOI{\tempurl}


\bibitem[Krizhevsky et~al\mbox{.}(2009)]%
        {krizhevsky2009learning}
\bibfield{author}{\bibinfo{person}{Alex Krizhevsky}, \bibinfo{person}{Geoffrey
  Hinton}, {et~al\mbox{.}}} \bibinfo{year}{2009}\natexlab{}.
\newblock \bibinfo{booktitle}{\emph{Learning multiple layers of features from
  tiny images}}.
\newblock \bibinfo{type}{{T}echnical {R}eport}.
  \bibinfo{institution}{University of Toronto}, \bibinfo{address}{Toronto, ON
  M5S, Canada}.
\newblock


\bibitem[Krizhevsky et~al\mbox{.}(2012)]%
        {krizhevsky2012imagenet}
\bibfield{author}{\bibinfo{person}{Alex Krizhevsky}, \bibinfo{person}{Ilya
  Sutskever}, {and} \bibinfo{person}{Geoffrey~E. Hinton}.}
  \bibinfo{year}{2012}\natexlab{}.
\newblock \showarticletitle{ImageNet Classification with Deep Convolutional
  Neural Networks}. In \bibinfo{booktitle}{\emph{Advances in Neural Information
  Processing Systems 25: 26th Annual Conference on Neural Information
  Processing Systems 2012}} (Lake Tahoe, Nevada, USA, December 3-6).
  \bibinfo{publisher}{Curran Associates, Inc.}, \bibinfo{address}{New York, NY,
  USA}, \bibinfo{pages}{1106--1114}.
\newblock
\urldef\tempurl%
\url{https://proceedings.neurips.cc/paper/2012/hash/c399862d3b9d6b76c8436e924a68c45b-Abstract.html}
\showURL{%
\tempurl}


\bibitem[LeCun et~al\mbox{.}(1998)]%
        {lecun1998gradient}
\bibfield{author}{\bibinfo{person}{Yann LeCun}, \bibinfo{person}{L{\'e}on
  Bottou}, \bibinfo{person}{Yoshua Bengio}, {and} \bibinfo{person}{Patrick
  Haffner}.} \bibinfo{year}{1998}\natexlab{}.
\newblock \showarticletitle{Gradient-based learning applied to document
  recognition}.
\newblock \bibinfo{journal}{\emph{Proc. IEEE}} \bibinfo{volume}{86},
  \bibinfo{number}{11} (\bibinfo{year}{1998}), \bibinfo{pages}{2278--2324}.
\newblock


\bibitem[Li et~al\mbox{.}(2020)]%
        {li2020review}
\bibfield{author}{\bibinfo{person}{Li Li}, \bibinfo{person}{Yuxi Fan},
  \bibinfo{person}{Mike Tse}, {and} \bibinfo{person}{Kuo-Yi Lin}.}
  \bibinfo{year}{2020}\natexlab{}.
\newblock \showarticletitle{A review of applications in federated learning}.
\newblock \bibinfo{journal}{\emph{Computers \& Industrial Engineering}}
  \bibinfo{volume}{149} (\bibinfo{year}{2020}), \bibinfo{pages}{106854}.
\newblock


\bibitem[Ligeng~Zhu(2019)]%
        {github:DRIA_github}
\bibfield{author}{\bibinfo{person}{Song~Han Ligeng~Zhu, Zhijian~Liu}.}
  \bibinfo{year}{2019}\natexlab{}.
\newblock \bibinfo{booktitle}{\emph{Deep Leakage From Gradients
  implementation}}.
\newblock github.
\newblock
\urldef\tempurl%
\url{https://github.com/mit-han-lab/dlg}
\showURL{%
Retrieved October 8,2022 from \tempurl}


\bibitem[Ling et~al\mbox{.}(2003)]%
        {ling2003auc}
\bibfield{author}{\bibinfo{person}{Charles~X. Ling}, \bibinfo{person}{Jin
  Huang}, {and} \bibinfo{person}{Harry Zhang}.}
  \bibinfo{year}{2003}\natexlab{}.
\newblock \showarticletitle{{AUC:} a Statistically Consistent and more
  Discriminating Measure than Accuracy}. In \bibinfo{booktitle}{\emph{IJCAI-03,
  Proceedings of the Eighteenth International Joint Conference on Artificial
  Intelligence}} (Acapulco, Mexico, August 9-15),
  \bibfield{editor}{\bibinfo{person}{Georg Gottlob} {and} \bibinfo{person}{Toby
  Walsh}} (Eds.). \bibinfo{publisher}{Morgan Kaufmann}, \bibinfo{address}{San
  Francisco, CA, USA}, \bibinfo{pages}{519--526}.
\newblock
\urldef\tempurl%
\url{http://ijcai.org/Proceedings/03/Papers/077.pdf}
\showURL{%
\tempurl}


\bibitem[Liu and Nocedal(1989)]%
        {liu1989limited}
\bibfield{author}{\bibinfo{person}{Dong~C Liu} {and} \bibinfo{person}{Jorge
  Nocedal}.} \bibinfo{year}{1989}\natexlab{}.
\newblock \showarticletitle{On the limited memory BFGS method for large scale
  optimization}.
\newblock \bibinfo{journal}{\emph{Mathematical programming}}
  \bibinfo{volume}{45}, \bibinfo{number}{1} (\bibinfo{year}{1989}),
  \bibinfo{pages}{503--528}.
\newblock


\bibitem[Melis et~al\mbox{.}(2019)]%
        {melis2019exploiting}
\bibfield{author}{\bibinfo{person}{Luca Melis}, \bibinfo{person}{Congzheng
  Song}, \bibinfo{person}{Emiliano~De Cristofaro}, {and}
  \bibinfo{person}{Vitaly Shmatikov}.} \bibinfo{year}{2019}\natexlab{}.
\newblock \showarticletitle{Exploiting Unintended Feature Leakage in
  Collaborative Learning}. In \bibinfo{booktitle}{\emph{2019 {IEEE} Symposium
  on Security and Privacy}} (San Francisco, CA, USA, May 19-23).
  \bibinfo{publisher}{IEEE}, \bibinfo{address}{Manhattan, NY, USA},
  \bibinfo{pages}{691--706}.
\newblock
\urldef\tempurl%
\url{https://doi.org/10.1109/SP.2019.00029}
\showDOI{\tempurl}


\bibitem[Mo et~al\mbox{.}(2022)]%
        {mo2022ppfl}
\bibfield{author}{\bibinfo{person}{Fan Mo}, \bibinfo{person}{Hamed Haddadi},
  \bibinfo{person}{Kleomenis Katevas}, \bibinfo{person}{Eduard Marin},
  \bibinfo{person}{Diego Perino}, {and} \bibinfo{person}{Nicolas Kourtellis}.}
  \bibinfo{year}{2022}\natexlab{}.
\newblock \showarticletitle{PPFL: Enhancing Privacy in Federated Learning with
  Confidential Computing}.
\newblock \bibinfo{journal}{\emph{GetMobile: Mobile Computing and
  Communications}} \bibinfo{volume}{25}, \bibinfo{number}{4}
  (\bibinfo{year}{2022}), \bibinfo{pages}{35--38}.
\newblock


\bibitem[Mo et~al\mbox{.}(2020)]%
        {mo2020darknetz}
\bibfield{author}{\bibinfo{person}{Fan Mo}, \bibinfo{person}{Ali~Shahin
  Shamsabadi}, \bibinfo{person}{Kleomenis Katevas}, \bibinfo{person}{Soteris
  Demetriou}, \bibinfo{person}{Ilias Leontiadis}, \bibinfo{person}{Andrea
  Cavallaro}, {and} \bibinfo{person}{Hamed Haddadi}.}
  \bibinfo{year}{2020}\natexlab{}.
\newblock \showarticletitle{DarkneTZ: Towards Model Privacy at the Edge Using
  Trusted Execution Environments}. In \bibinfo{booktitle}{\emph{Proceedings of
  the 18th International Conference on Mobile Systems, Applications, and
  Services}} (Toronto, Ontario, Canada). \bibinfo{publisher}{ACM},
  \bibinfo{address}{New York, NY, USA}, \bibinfo{pages}{161–174}.
\newblock
\urldef\tempurl%
\url{https://doi.org/10.1145/3386901.3388946}
\showDOI{\tempurl}


\bibitem[Ménétrey et~al\mbox{.}(2022)]%
        {watz}
\bibfield{author}{\bibinfo{person}{Jämes Ménétrey}, \bibinfo{person}{Marcelo
  Pasin}, \bibinfo{person}{Pascal Felber}, {and} \bibinfo{person}{Valerio
  Schiavoni}.} \bibinfo{year}{2022}\natexlab{}.
\newblock \showarticletitle{WaTZ: A Trusted WebAssembly Runtime Environment
  with Remote Attestation for TrustZone}. In \bibinfo{booktitle}{\emph{2022
  IEEE 42nd IEEE International Conference on Distributed Computing Systems
  (ICDCS)}} (Bologna, Italy, July 10-13). \bibinfo{publisher}{IEEE},
  \bibinfo{address}{Manhattan, NY, USA}, \bibinfo{pages}{1177--1189}.
\newblock
\urldef\tempurl%
\url{https://doi.org/10.1109/ICDCS54860.2022.00116}
\showDOI{\tempurl}


\bibitem[Nasr et~al\mbox{.}(2019)]%
        {nasr2019comprehensive}
\bibfield{author}{\bibinfo{person}{Milad Nasr}, \bibinfo{person}{Reza Shokri},
  {and} \bibinfo{person}{Amir Houmansadr}.} \bibinfo{year}{2019}\natexlab{}.
\newblock \showarticletitle{Comprehensive Privacy Analysis of Deep Learning:
  Passive and Active White-box Inference Attacks against Centralized and
  Federated Learning}. In \bibinfo{booktitle}{\emph{2019 {IEEE} Symposium on
  Security and Privacy}} (San Francisco, CA, USA). \bibinfo{publisher}{{IEEE}},
  \bibinfo{address}{Manhattan, NY, USA}, \bibinfo{pages}{739--753}.
\newblock
\urldef\tempurl%
\url{https://doi.org/10.1109/SP.2019.00065}
\showDOI{\tempurl}


\bibitem[Peltarion(2022)]%
        {web:crossentropy}
\bibfield{author}{\bibinfo{person}{Peltarion}.}
  \bibinfo{year}{2022}\natexlab{}.
\newblock \bibinfo{booktitle}{\emph{Categorical crossentropy}}.
\newblock
\urldef\tempurl%
\url{https://peltarion.com/knowledge-center/documentation/modeling-view/build-an-ai-model/loss-functions/categorical-crossentropy}
\showURL{%
Retrieved October 8, 2022 from \tempurl}


\bibitem[Pinto and Santos(2019)]%
        {pinto2019demystifying}
\bibfield{author}{\bibinfo{person}{Sandro Pinto} {and} \bibinfo{person}{Nuno
  Santos}.} \bibinfo{year}{2019}\natexlab{}.
\newblock \showarticletitle{Demystifying arm trustzone: A comprehensive
  survey}.
\newblock \bibinfo{journal}{\emph{ACM Computing Surveys (CSUR)}}
  \bibinfo{volume}{51}, \bibinfo{number}{6} (\bibinfo{year}{2019}),
  \bibinfo{pages}{1--36}.
\newblock


\bibitem[Redmon(2016)]%
        {darknet}
\bibfield{author}{\bibinfo{person}{Joseph Redmon}.}
  \bibinfo{year}{2013--2016}\natexlab{}.
\newblock \bibinfo{booktitle}{\emph{Darknet: Open Source Neural Networks in
  C}}.
\newblock
\urldef\tempurl%
\url{http://pjreddie.com/darknet/}
\showURL{%
Retrieved October 8, 2022 from \tempurl}


\bibitem[Rieke et~al\mbox{.}(2020)]%
        {rieke2020future}
\bibfield{author}{\bibinfo{person}{Nicola Rieke}, \bibinfo{person}{Jonny
  Hancox}, \bibinfo{person}{Wenqi Li}, \bibinfo{person}{Fausto Milletari},
  \bibinfo{person}{Holger~R Roth}, \bibinfo{person}{Shadi Albarqouni},
  \bibinfo{person}{Spyridon Bakas}, \bibinfo{person}{Mathieu~N Galtier},
  \bibinfo{person}{Bennett~A Landman}, \bibinfo{person}{Klaus Maier-Hein},
  {et~al\mbox{.}}} \bibinfo{year}{2020}\natexlab{}.
\newblock \showarticletitle{The future of digital health with federated
  learning}.
\newblock \bibinfo{journal}{\emph{NPJ digital medicine}} \bibinfo{volume}{3},
  \bibinfo{number}{1} (\bibinfo{year}{2020}), \bibinfo{pages}{1--7}.
\newblock


\bibitem[Rodr{\'\i}guez(2017)]%
        {rodriguez2017evolution}
\bibfield{author}{\bibinfo{person}{Ricardo~J Rodr{\'\i}guez}.}
  \bibinfo{year}{2017}\natexlab{}.
\newblock \showarticletitle{Evolution and characterization of point-of-sale RAM
  scraping malware}.
\newblock \bibinfo{journal}{\emph{Journal of Computer Virology and Hacking
  Techniques}} \bibinfo{volume}{13}, \bibinfo{number}{3}
  (\bibinfo{year}{2017}), \bibinfo{pages}{179--192}.
\newblock


\bibitem[Sabt et~al\mbox{.}(2015)]%
        {7345265}
\bibfield{author}{\bibinfo{person}{Mohamed Sabt}, \bibinfo{person}{Mohammed
  Achemlal}, {and} \bibinfo{person}{Abdelmadjid Bouabdallah}.}
  \bibinfo{year}{2015}\natexlab{}.
\newblock \showarticletitle{Trusted Execution Environment: What It is, and What
  It is Not}. In \bibinfo{booktitle}{\emph{{IEEE} TrustCom/BigDataSE/ISPA}}
  (Helsinki, Finland, August 20-22). \bibinfo{publisher}{IEEE},
  \bibinfo{address}{Manhattan, NY, USA}, \bibinfo{pages}{57--64}.
\newblock
\urldef\tempurl%
\url{https://doi.org/10.1109/Trustcom.2015.357}
\showDOI{\tempurl}


\bibitem[Song(2018)]%
        {github:DPIA_github}
\bibfield{author}{\bibinfo{person}{Congzheng Song}.}
  \bibinfo{year}{2018}\natexlab{}.
\newblock \bibinfo{booktitle}{\emph{Property Inference in Collaborative ML
  implementation}}.
\newblock github.
\newblock
\urldef\tempurl%
\url{https://github.com/csong27/property-inference-collaborative-ml}
\showURL{%
Retrieved October 8, 2022 from \tempurl}


\bibitem[Tram{\`{e}}r and Boneh(2019)]%
        {tramer2018slalom}
\bibfield{author}{\bibinfo{person}{Florian Tram{\`{e}}r} {and}
  \bibinfo{person}{Dan Boneh}.} \bibinfo{year}{2019}\natexlab{}.
\newblock \showarticletitle{Slalom: Fast, Verifiable and Private Execution of
  Neural Networks in Trusted Hardware}. In \bibinfo{booktitle}{\emph{7th
  International Conference on Learning Representations, {ICLR} 2019}} (New
  Orleans, LA, USA, May 6-9). \bibinfo{publisher}{OpenReview.net},
  \bibinfo{address}{Online website}, \bibinfo{numpages}{19}~pages.
\newblock
\urldef\tempurl%
\url{https://openreview.net/forum?id=rJVorjCcKQ}
\showURL{%
\tempurl}


\bibitem[TrustedFirmware(2021a)]%
        {doc:optee}
\bibfield{author}{\bibinfo{person}{TrustedFirmware}.}
  \bibinfo{year}{2021}\natexlab{a}.
\newblock \bibinfo{booktitle}{\emph{OP-TEE Documentation}}.
\newblock TrustedFirmware.
\newblock
\urldef\tempurl%
\url{https://optee.readthedocs.io/en/latest/}
\showURL{%
Retrieved October 8, 2022 from \tempurl}


\bibitem[TrustedFirmware(2021b)]%
        {web:securestorage}
\bibfield{author}{\bibinfo{person}{TrustedFirmware}.}
  \bibinfo{year}{2021}\natexlab{b}.
\newblock \bibinfo{booktitle}{\emph{Secure storage}}.
\newblock TrustedFirmware.
\newblock
\urldef\tempurl%
\url{https://optee.readthedocs.io/en/latest/architecture/secure_storage.html}
\showURL{%
Retrieved October 8, 2022 from \tempurl}


\bibitem[Trustonic(2021)]%
        {web:trustonic}
\bibfield{author}{\bibinfo{person}{Trustonic}.}
  \bibinfo{year}{2021}\natexlab{}.
\newblock \bibinfo{booktitle}{\emph{Kinibi-520a: The latest Trustonic Trusted
  Execution Environment (TEE)}}.
\newblock
\urldef\tempurl%
\url{https://www.trustonic.com/technical-articles/kinibi-520a-the-latest-trusted-execution-environment-tee/}
\showURL{%
Retrieved October 8, 2022 from \tempurl}


\bibitem[Umar and Mayes(2017)]%
        {umar2017trusted}
\bibfield{author}{\bibinfo{person}{Assad Umar} {and} \bibinfo{person}{Keith
  Mayes}.} \bibinfo{year}{2017}\natexlab{}.
\newblock \showarticletitle{Trusted Execution Environment and Host Card
  Emulation}.
\newblock In \bibinfo{booktitle}{\emph{Smart Cards, Tokens, Security and
  Applications}}. \bibinfo{publisher}{Springer}, \bibinfo{address}{New York,
  NY, USA}, \bibinfo{pages}{497--519}.
\newblock


\bibitem[Wei et~al\mbox{.}(2020)]%
        {wei2020federated}
\bibfield{author}{\bibinfo{person}{Kang Wei}, \bibinfo{person}{Jun Li},
  \bibinfo{person}{Ming Ding}, \bibinfo{person}{Chuan Ma},
  \bibinfo{person}{Howard~H Yang}, \bibinfo{person}{Farhad Farokhi},
  \bibinfo{person}{Shi Jin}, \bibinfo{person}{Tony~QS Quek}, {and}
  \bibinfo{person}{H~Vincent Poor}.} \bibinfo{year}{2020}\natexlab{}.
\newblock \showarticletitle{Federated learning with differential privacy:
  Algorithms and performance analysis}.
\newblock \bibinfo{journal}{\emph{IEEE Transactions on Information Forensics
  and Security}}  \bibinfo{volume}{15} (\bibinfo{year}{2020}),
  \bibinfo{pages}{3454--3469}.
\newblock


\bibitem[Yang et~al\mbox{.}(2018)]%
        {yang2018applied}
\bibfield{author}{\bibinfo{person}{Timothy Yang}, \bibinfo{person}{Galen
  Andrew}, \bibinfo{person}{Hubert Eichner}, \bibinfo{person}{Haicheng Sun},
  \bibinfo{person}{Wei Li}, \bibinfo{person}{Nicholas Kong},
  \bibinfo{person}{Daniel Ramage}, {and} \bibinfo{person}{Fran{\c{c}}oise
  Beaufays}.} \bibinfo{year}{2018}\natexlab{}.
\newblock \showarticletitle{Applied Federated Learning: Improving Google
  Keyboard Query Suggestions}.
\newblock  (\bibinfo{year}{2018}).
\newblock
\urldef\tempurl%
\url{https://doi.org/10.48550/arXiv.1812.02903}
\showDOI{\tempurl}
\showeprint[arXiv]{1812.02903}


\bibitem[Yin et~al\mbox{.}(2021)]%
        {yin2021see}
\bibfield{author}{\bibinfo{person}{Hongxu Yin}, \bibinfo{person}{Arun Mallya},
  \bibinfo{person}{Arash Vahdat}, \bibinfo{person}{Jose~M. Alvarez},
  \bibinfo{person}{Jan Kautz}, {and} \bibinfo{person}{Pavlo Molchanov}.}
  \bibinfo{year}{2021}\natexlab{}.
\newblock \showarticletitle{See Through Gradients: Image Batch Recovery via
  GradInversion}. In \bibinfo{booktitle}{\emph{{IEEE} Conference on Computer
  Vision and Pattern Recognition, {CVPR} 2021}} (Virtual Event, June 19-25).
  \bibinfo{publisher}{Computer Vision Foundation / {IEEE}},
  \bibinfo{address}{Manhattan, NY, USA}, \bibinfo{pages}{16337--16346}.
\newblock
\urldef\tempurl%
\url{https://doi.org/10.1109/CVPR46437.2021.01607}
\showDOI{\tempurl}


\bibitem[Zhang et~al\mbox{.}(2020)]%
        {zhang2020batchcrypt}
\bibfield{author}{\bibinfo{person}{Chengliang Zhang}, \bibinfo{person}{Suyi
  Li}, \bibinfo{person}{Junzhe Xia}, \bibinfo{person}{Wei Wang},
  \bibinfo{person}{Feng Yan}, {and} \bibinfo{person}{Yang Liu}.}
  \bibinfo{year}{2020}\natexlab{}.
\newblock \showarticletitle{BatchCrypt: Efficient Homomorphic Encryption for
  Cross-Silo Federated Learning}. In \bibinfo{booktitle}{\emph{Proceedings of
  the 2020 USENIX Conference on Usenix Annual Technical Conference}} (Online
  event, July 15-17). \bibinfo{publisher}{USENIX Association},
  \bibinfo{address}{Berkeley, CA, USA}, \bibinfo{pages}{493--506}.
\newblock


\bibitem[Zhang et~al\mbox{.}(2021)]%
        {zhang2021citadel}
\bibfield{author}{\bibinfo{person}{Chengliang Zhang}, \bibinfo{person}{Junzhe
  Xia}, \bibinfo{person}{Baichen Yang}, \bibinfo{person}{Huancheng Puyang},
  \bibinfo{person}{Wei Wang}, \bibinfo{person}{Ruichuan Chen},
  \bibinfo{person}{Istemi~Ekin Akkus}, \bibinfo{person}{Paarijaat Aditya},
  {and} \bibinfo{person}{Feng Yan}.} \bibinfo{year}{2021}\natexlab{}.
\newblock \showarticletitle{Citadel: Protecting Data Privacy and Model
  Confidentiality for Collaborative Learning}. In
  \bibinfo{booktitle}{\emph{Proceedings of the ACM Symposium on Cloud
  Computing}} (Seattle, WA, USA, November 1-4). \bibinfo{publisher}{ACM},
  \bibinfo{address}{New York, NY, USA}, \bibinfo{pages}{546–561}.
\newblock
\urldef\tempurl%
\url{https://doi.org/10.1145/3472883.3486998}
\showDOI{\tempurl}


\bibitem[Zhao et~al\mbox{.}(2022)]%
        {sear}
\bibfield{author}{\bibinfo{person}{Lingchen Zhao}, \bibinfo{person}{Jianlin
  Jiang}, \bibinfo{person}{Bo Feng}, \bibinfo{person}{Qian Wang},
  \bibinfo{person}{Chao Shen}, {and} \bibinfo{person}{Qi Li}.}
  \bibinfo{year}{2022}\natexlab{}.
\newblock \showarticletitle{SEAR: Secure and Efficient Aggregation for
  Byzantine-Robust Federated Learning}.
\newblock \bibinfo{journal}{\emph{IEEE Transactions on Dependable and Secure
  Computing}} \bibinfo{volume}{19}, \bibinfo{number}{5} (\bibinfo{year}{2022}),
  \bibinfo{pages}{3329--3342}.
\newblock
\urldef\tempurl%
\url{https://doi.org/10.1109/TDSC.2021.3093711}
\showDOI{\tempurl}


\bibitem[Zhao et~al\mbox{.}(2014)]%
        {zhao2014providing}
\bibfield{author}{\bibinfo{person}{Shijun Zhao}, \bibinfo{person}{Qianying
  Zhang}, \bibinfo{person}{Guangyao Hu}, \bibinfo{person}{Yu Qin}, {and}
  \bibinfo{person}{Dengguo Feng}.} \bibinfo{year}{2014}\natexlab{}.
\newblock \showarticletitle{Providing Root of Trust for ARM TrustZone Using
  On-Chip SRAM}. In \bibinfo{booktitle}{\emph{Proceedings of the 4th
  International Workshop on Trustworthy Embedded Devices, TrustED '14}}
  (Scottsdale, Arizona, USA, November 3). \bibinfo{publisher}{ACM},
  \bibinfo{address}{New York, NY, USA}, \bibinfo{pages}{25–36}.
\newblock
\urldef\tempurl%
\url{https://doi.org/10.1145/2666141.2666145}
\showDOI{\tempurl}


\bibitem[Zhu et~al\mbox{.}(2019)]%
        {zhu2019deep}
\bibfield{author}{\bibinfo{person}{Ligeng Zhu}, \bibinfo{person}{Zhijian Liu},
  {and} \bibinfo{person}{Song Han}.} \bibinfo{year}{2019}\natexlab{}.
\newblock \showarticletitle{Deep Leakage from Gradients}. In
  \bibinfo{booktitle}{\emph{Advances in Neural Information Processing Systems
  32: Annual Conference on Neural Information Processing Systems 2019}}
  (Vancouver, BC, Canada, December 8-14). \bibinfo{publisher}{Curran
  Associates, Inc.}, \bibinfo{address}{New York, NY, USA},
  \bibinfo{numpages}{11}~pages.
\newblock
\urldef\tempurl%
\url{https://doi.org/10.48550/arXiv.1906.08935}
\showDOI{\tempurl}


\end{thebibliography}
